\documentclass[%
 reprint,
floatfix,
superscriptaddress,
showpacs,preprintnumbers,
 amsmath,amssymb,
 aps,
pra,
]{revtex4-1}
\usepackage{xcolor}
\usepackage{times}
\usepackage{amssymb}
\usepackage[notrig]{physics}
\usepackage{graphicx}
\usepackage{dcolumn}
\usepackage{bm}
\usepackage{epstopdf}
\usepackage{lineno} 
\usepackage{hyperref}
\newcommand*\chem[1]{\ensuremath{\mathrm{#1}}}

\begin{document}
\title{\textbf{Charge, bond, and pair density wave orders in a strongly correlated system}}

\author{Anurag Banerjee}
\affiliation{Institut de Physique Th\'eorique, Universit\'{e} Paris-Saclay, CEA, CNRS, F-91191 Gif-sur-Yvette, France}

\author{Catherine P\'epin}
\affiliation{Institut de Physique Th\'eorique, Universit\'{e} Paris-Saclay, CEA, CNRS, F-91191 Gif-sur-Yvette, France}

\author{Amit Ghosal}
\affiliation{Indian Institute of Science Education and Research Kolkata, Mohanpur, India-741246}

\begin{abstract}
The coexistence of multiple quasi-degenerate orders is the hallmark of the strongly correlated materials. Experiments often reveal several spatially modulated orders in the underdoped cuprates. This has come to the forefront with the possible detection of the pair density wave states. However, microscopic calculations often struggle to stabilize such spatially modulating orders as the ground state in the strong correlation limit. This work uses the $t-t^\prime-J$-model with an additional nearest-neighbor repulsion to stabilize spatially oscillating charge, bond, and pairing orders in the underdoped regime. We employ the standard Gutzwiller approach while treating the inhomogeneity for the spatial orders using the self-consistent Hartree-Fock-Bogoliubov methodology. Our calculations reveal that unidirectional bond density states coexisting with charge and pairing modulations can have lower energy than the uniform superconducting state over an extensive doping range. These modulating states vanish monotonically as the modulation wavevector becomes shorter with increased dopings. The finite momentum orders give way to a vestigial nematic phase upon increasing doping which only breaks the rotational symmetry of the system. The nematic order vanishes upon further increasing doping, and only uniform superconductivity survives. The spatial features of the ground state at each doping reveal multiple wavevectors, which potentially drives the incommensuration of charge orders. Interestingly, the spatially modulating states are absent when the strong correlations criteria are relaxed, suggesting that the removal of double occupancy aids the stabilization of density wave orders.
\end{abstract}
\maketitle
\section{Introduction}
\label{introduction}
The presence of multiple broken symmetry phases and competing orders has become the new paradigm for describing correlated quantum matter featuring complex phase diagram~\cite{keimer2017,sun2016dome,dagotto2005complexity,bounoua2020loop}. Given the abundance of rich phase diagrams in different materials suggests the intertwining of several orders~\cite{Fradkin:2015ch,Vestigial19_ARCMP}. In this scenario, a precursor state of all potentially degenerate states forms at a high energy scale which cannot be linked with one particular order. At low energies, slight energy differences stabilize one order over the other. 

A unique realization for intertwined orders is the vestigial or composite phase, where a higher-order component of the likely symmetry broken phase condenses although the individual components vanish~\cite{Fradkin:2015ch,Vestigial19_ARCMP}. Therefore, a vestigial state has a smaller subset of broken symmetries than the parent state. For instance, a nematic phase that breaks the rotational symmetry is a vestige of the primary checkerboard and stripe phase where both translation and rotational symmetries are broken~\cite{Nie13}. Recently, the sublattice phase-resolved study of electronic structure in Bi-based cuprates compound finds a vestigial nematic phase around $14-17\%$ hole doping~\cite{nematic19PNAS}. Moreover, the nematic order coexists with the charge density wave (CDW) orders at lower dopings~\cite{nematic19PNAS}.     

The signatures of charge modulation are routinely observed at moderate hole-doped materials in all cuprate families~\cite{Hoffman02,Wise08,frano2020charge}. Usually, the modulation wavevector, $\mathbf{Q}$, reduces as the doping increases~\cite{Comin14,daSilvaNeto14} except for the \chem{La_2CuO_4} where the wavevector increases with doping~\cite{Fink09,Fink11,Hucker11}. Moreover, detailed investigations have also revealed a bond-centered charge order with a primarily d-wave form factor~\cite{forgan2015,Comin:2015ca,Comin15a}. Furthermore, disorder severely affects the charge modulations such that the unidirectional stripe and the bidirectional checkerboard patterns become indistinguishable~\cite{Campi15}. The local spectroscopic probes indicate short-ranged domains with unidirectional charge modulation~\cite{Kohsaka07,Parker:2010if}, albeit quantum oscillation measurements favor the bidirectional picture~\cite{chan2016single}. 

Recently the study of the modulated orders in cuprates has gained impetus with the observations of pair density wave (PDW) state coexisting with usual $d$-wave superconductivity (SC) using Josephson scanning tunneling microscopy (STM)~\cite{Hamidian16,du2020imaging}. In the same regime, $d$-wave CDW oscillations are also detected\cite{du2020imaging}. The importance of detecting the PDW state is underlined by the several exotic proposals of the celebrated pseudogap state that hinges on the quantum disordering~\cite{Dai19} or fractionalization of the PDW state~\cite{Chakraborty19,Grandadam19}. However, it is extremely difficult to stabilize such PDW sates as the ground state of a microscopic Hamiltonian~\cite{Lee14,xu2019pair,berg2010pair,peng2021precursor,dash2021}.

The $tJ$-model and its parent Hubbard model at large repulsion $U$ are generally studied to extract the physics of spatially modulated states~\cite{Sachdev13,HubbardStripe17,tJPDW0}. The variational approaches on the two-dimensional (2D) $tJ$ model find a striking near-degeneracy among the unidirectional charge order with stripe SC, a PDW state with spin density wave component, and the uniform $d$-wave SC state~\cite{Corboz14}. A few recent calculations on the same model observe a pair density wave state with vanishing mean SC pairing energetically very close to the uniform d-wave SC~\cite{choubey2017,choubey2020}. Such quasi-degeneracy among different states suggests that a small perturbation to the $tJ$ model will favor one state over the other, thus giving rise to a wide variety of phenomenology.

Using commonly accepted minimal models to describe strongly correlated quantum matters, i.e., the $t-t^\prime-J$ model with an additional nearest-neighbor repulsion, we study the interplay of unconventional superconductivity with spatially modulated charge-ordered states in the ground state. Early investigations on such model manifest promising signatures of density wave orders~\cite{Allais14c,Allais14b,spalek2017,sau2014mean,ZegrodnikPRB}. A careful analysis of the variation of the ordering wavevector with doping, however, is still missing.
In addition, the real-space pattern of the charge, bond order, and the possibility of finding spatially modulated pairing amplitude (similar to the Fulde-Ferrell-Larkin-Ovchinnikov phase {\it but} in the absence of magnetic field) need to be studied. Furthermore, the mechanism of the demise of modulated orders with increasing doping remains unexplored. In this work, we focus on the unidirectional charge orders that are routinely observed in the underdoped cuprates~\cite{Kohsaka07,Parker:2010if,zhao2019charge}. Furthermore, we neglect the magnetically ordered states to keep the numerical calculation tractable. In the next section, we describe the model and the method in detail.

\section{Model and Methods}
We work with the standard $tJ$-model on a two-dimensional square lattice with all the double occupancy removed, with an additional nearest neighbor repulsion. The model is given by
\begin{align}
    \mathcal{H}=\sum_{i,j,\sigma} &\left(t_{ij} \tilde{c}^{\dagger}_{i \sigma} \tilde{c}_{j \sigma} + h.c. \right) + J\sum_{\langle i,j \rangle}  \left(\tilde{\mathbf{S}}_i \cdot \tilde{\mathbf{S}}_j -\frac{1}{4}\tilde{n}_i \tilde{n}_j\right)  \nonumber \\
    &+ W \sum_{\langle i,j \rangle,\sigma,\sigma^\prime}\hat{n}_{i \sigma} \hat{n}_{j \sigma \prime} - \mu \sum_{i,\sigma} \hat{n}_{i,\sigma}.
    \label{eq:tj}
    \end{align}
Here, $t_{ij}=-t$ if $i,j$ are nearest neighbors and $t_{ij}=t^{\prime}$ if $i,j$ are next nearest neighbor sites. We fix the $t=1$ and $t^{\prime}=0.25t$, and all the energy scales are in the units of $t$. To compare with the experimental energy scales in cuprates, one can use $t\approx300$ meV.

The $tJ$ Hamiltonian is an effective low-energy model obtained from a Schrieffer-Wolf transformation of the Hubbard model for large on-site repulsion $U$~\cite{chao1977kinetic}. The $\mathbf{S}_i$ are the electron spin operator and the exchange interaction $J\approx(4t^2)/U$, is set to $J=0.3t$ in this work. Furthermore, $W$ is the nearest neighbor repulsion term, which we fix to $W=0.6t$ for the rest of the analysis. 

The local number operators for a particular spin is given by $\hat{n}_{i,\sigma}=c^\dagger_{i,\sigma} c_{i,\sigma}$ and the local density is the expectation value of the number operator. The `tilde' appearing on the creation (annihilation) operators at site $i$ with spin $\sigma=\uparrow,\downarrow$, $c^\dagger_{i,\sigma}$ ($c_{i,\sigma}$) in Eq.~(\ref{eq:tj}) is explained below. The local density is thus given by ${\rho_i=\sum_\sigma \langle c^\dagger_{i\sigma} c_{i\sigma} \rangle}$. $\mu$ is the chemical potential that fixes the average density of carriers $\rho$ in the system. The spatial average of the local density gives the average density of electrons $\rho$. The (under) doping level from the half-filling is defined via ${\delta=(1-\rho)}$. 

Strong on-site repulsions reduce the Hilbert space of the $tJ$-model by eliminating all double occupancies. The `tilde' on operators appearing in the Eq.~(\ref{eq:tj}) signifies that they operate on a truncated Hilbert space which projects out double occupation on any site due to strong repulsive interactions. The operators without `tilde' live in the usual unrestricted Hilbert space.
\begin{align}
    \tilde{c}_{i \sigma}=c_{i\sigma} (1-\hat{n}_{i\overline{\sigma}}),
\end{align}
where $c$ is the normal annihilation operator. A straightforward implementation of such a restriction is given by the Gutzwiller approximation, which transforms the strongly interacting problem to an unrestricted Hilbert space, at the expense of introducing ``Gutzwiller factors", which renormalize all operators~\cite{fazekas1999lecture,fukushima2008grand,ko2007extended}.
\begin{align}
    \langle \tilde{c}^\dagger_{i\sigma} \tilde{c}_{j\sigma} \rangle &\approx g^t_{ij}\langle c^\dagger_{i\sigma} c_{j\sigma} \rangle_0,\\
     \langle \tilde{\mathbf{S}}_i.\tilde{\mathbf{S}}_{j} \rangle &\approx g^{xy}_{ij} \langle \mathbf{S}_i. \mathbf{S}_j \rangle_0,\\
     \langle n_{i\sigma} n_{j\sigma} \rangle &\approx \langle n_{i\sigma} n_{j\sigma} \rangle_0,
\end{align}
where $\langle \dots \rangle$ is the expectation in the restricted Hilbert space whereas  $\langle \dots \rangle_0$ is the expectation of operators in the unrestricted Hilbert space. The Gutzwiller factors in the absence of any magnetic orders are given by
\begin{align}
    g^t_{ij}&=\sqrt{\frac{4 x_i x_j}{(1+x_i)(1+x_j)}},\\
    g^{xy}_{ij}&=\frac{4}{(1+x_i)(1+x_j)},
\end{align}
where $x_i=1-\rho_i$. After removing the double-occupancy, we perform the mean field decomposition of the interaction terms in Hartree, Fock and Cooper channels as given below,
\begin{align}
    \sum_{\sigma}\langle c^\dagger_{i\sigma} c_{i\sigma} \rangle_0 &= \rho_i,\\
    \langle c^\dagger_{i\sigma} c_{j\sigma} \rangle_0 &= \chi_{ij},\\
    \langle c_{j\downarrow} c_{i \uparrow} + c_{i\downarrow} c_{j\uparrow}\rangle_0 &=\Delta_{ij}.
\end{align}
Note that we are not allowing for any magnetic orderings in this system by focusing on superconducting with charge and bond density wave (BDW) orders only. Therefore, we eliminate the spin-flip Fock terms from the mean-field Hamiltonian, to preserve the spin rotational symmetry. To perform the mean field analysis we closely follow Ref.~\cite{choubey2017,Chakraborty:2014iq,garg2008strong}. The explicit expression for the mean-field Hamiltonian is given by
\begin{align}
    \mathcal{H}_{MF}=&\sum_{i,a,\sigma} \left( t g^t_{i,i+a}- W^{FS}_{i,i+a}\right) c^\dagger_{i,\sigma} c_{i+a,\sigma} \nonumber \\&+ \sum_{i,b,\sigma} t^\prime g^t_{i,i+b} c^\dagger_{i,\sigma} c_{i+b,\sigma} +\sum_{i,\sigma} (-\mu+\mu^{HS}_i) \hat{n}_{i\sigma} \nonumber\\ &+ \sum_{i,a} \left[ G^{i,a}_1 \Delta_{i,i+a} \left( c^\dagger_{i,\uparrow} c^\dagger_{i+a,\downarrow}+ c^\dagger_{i+a,\uparrow} c^\dagger_{i,\downarrow}\right)+h.c.\right].
\end{align}
Here, the sum over $\mathbf{a}=\pm \hat{x},\pm \hat{y}$ are the nearest neighbor vectors and $\mathbf{b}=\pm(\hat{x} \pm \hat{y})$ are the next neighbor vector. The other variables are given by,
\begin{align}
    W^{FS}_{i,a}&=\frac{J}{2} \left(\frac{3 g^{xy}_{i,i+a}}{2} -\frac{1}{2} \right) \chi_{i,i+a} +W \chi_{i,i+a},\\
    G^{i,a}_1&=-\frac{J}{4} \left(\frac{3 g^{xy}_{i,i+a}+1}{4} \right), \\
    \mu^{HS}_i&=-4t \sum_{a,\sigma} \frac{\partial g^{t}_{i,i+a}}{\partial n_i} \chi_{i,i+a} +4t^\prime \sum_{b,\sigma} \frac{\partial g^{t}_{i,i+b}}{\partial n_i} \chi_{i,i+b}\nonumber \\ &-\frac{3J}{2} \sum_{a,\sigma} \frac{\partial g^{xy}_{i,i+a}}{\partial n_i} \left( \frac{\Delta^2_{i,i+a}}{4} + \chi^2_{i,i+a}\right)\nonumber \\
    &-\frac{J-2W}{4}\sum_{i,a,\sigma} \rho_{i+a}.
\end{align}
Next, we diagonalize the Hamiltonian using Bogoliubov-deGennes(BdG) transformation\cite{PdGBook,Ghosal98,Ghosal01}, 
\begin{align}
    c_{i\sigma}=\sum_n \left(\gamma_{n\sigma} u_{i,n} -\sigma \gamma^\dagger_{n,-\sigma} v^*_{i,n} \right),
\end{align}
where $\gamma^\dagger (\gamma)$ are Bogoliubov creation (annihilation) operator.  The resulting eigensystem is then solved self-consistently for all the local Hartree, Fock and Bogoliubov order parameters and $\mu$. If there are $N$ lattice sites, then there are $(5N+1)$ variables to be solved self-consistently. We initialize the SC pairing $\Delta_{ij}$ and and the bond density $\chi_{ij}$, and local density to be the modulating around a mean value with a desired wavevector $\mathbf{Q}=(1/\lambda) 2\pi/a_0 \hat{x}$,. Here $\lambda$ is the period of oscillations and $a_0$ is the lattice-spacing. We focus on the unidirectional density wave orders in this paper. The initial guess is given by
\begin{align}
    \Delta_{i,i+a}=\Delta_0 + \Delta_Q \cos(\mathbf{Q}.r_i).
    \label{eq:guess}
\end{align}
$\Delta_0$ is the guess mean SC pairing amplitude, and $\Delta_Q$ is the initial guess for SC pairing modulation amplitude. We initialize the bond and local density similarly and use such a trail solution to achieve self-consistency. We perform further analysis on those solutions for the cases we achieve self-consistency with the desired wavevector.

One can evaluate the self-consistent d-wave and extended s-wave components from the SC pairing $\Delta_{ij}$ by using the following definition
\begin{align}
    \Delta_d(i)=\frac{1}{4}\left(\Delta_{i,i+\hat{x}} + \Delta_{i,i-\hat{x}} - \Delta_{i,i+\hat{y}} - \Delta_{i,i-\hat{y}}   \right),\\
    \Delta_s(i)=\frac{1}{4}\left(\Delta_{i,i+\hat{x}} + \Delta_{i,i-\hat{x}} + \Delta_{i,i+\hat{y}} + \Delta_{i,i-\hat{y}}   \right).
\end{align}
Similar definition can be made for the d-wave and extended s-wave component for the bond density order $\chi_{ij}$.
\begin{align}
    \chi_d(i)=\frac{1}{4}\left(\chi_{i,i+\hat{x}} + \chi_{i,i-\hat{x}} - \chi_{i,i+\hat{y}} - \chi_{i,i-\hat{y}}   \right), \label{eq:chid}\\
    \chi_s(i)=\frac{1}{4}\left(\chi_{i,i+\hat{x}} + \chi_{i,i-\hat{x}} + \chi_{i,i+\hat{y}} + \chi_{i,i-\hat{y}}   \right). \label{eq:chis}
\end{align}
To extract the finite-momentum component of the charge, bond and pairing, we first calculate the Fourier components of the variation of an ordering. For example, the Fourier component for the local density, is given by
\begin{align}
    \tilde{\rho}({\mathbf{q}})=\frac{1}{N} \sum_{i} \exp({i \mathbf{q}.\mathbf{r}_i}) (\rho_i-\rho).
    \label{eq:chi_rho}
\end{align}
Note that we remove the average density from the local density to consider only the oscillating part instead of the uniform part. The principal peak from the Fourier component gives us the modulation wavevector $Q$ for the desired quantity. The modulation amplitude from the local density is evaluated by the strength of $\tilde{\rho}(q)$ at this wavevector such that $\chi^Q_\rho\equiv\tilde{\rho}(\mathbf{Q})$. Similarly, we can define the other modulation amplitude by using the Fourier transform of the respective order parameters, as
\begin{align}
    \tilde{\chi}_d({\mathbf{q}})&=\frac{1}{N} \sum_{i} \exp({i \mathbf{q}.\mathbf{r}_i}) (\chi_d(i)-\chi_d),
    \label{eq:chi_d}\\
    \tilde{\chi}_s({\mathbf{q}})&=\frac{1}{N} \sum_{i} \exp({i \mathbf{q}.\mathbf{r}_i}) (\chi_s(i)-\chi_s),
    \label{eq:chi_s}\\
    \tilde{\Delta}_d({\mathbf{q}})&=\frac{1}{N} \sum_{i} \exp({i \mathbf{q}.\mathbf{r}_i}) (\Delta_d(i)-\Delta_d),\label{eq:delta_d}\\
    \tilde{\Delta}_s({\mathbf{q}})&=\frac{1}{N} \sum_{i} \exp({i \mathbf{q}.\mathbf{r}_i}) (\Delta_s(i)-\Delta_s).
    \label{eq:delta_s}
\end{align}
Here $\chi_d$ are the spatially averaged component d-wave bond density. The finite-momentum component at the dominant wavevector is again defined as $\chi^Q_d\equiv\tilde{\chi}_d(\mathbf{Q})$. Similar definitions are also used for the other variables.

We can also track the broken rotational symmetry of the bond density and cooper pair density operators by using~\cite{Nie13}
\begin{align}
    \mathcal{N}_{\chi} &=\frac{1}{2} \sqrt{\langle \chi_x^2 \rangle -\langle \chi_y^2 \rangle }\\
    \mathcal{N}_{\Delta} &=\frac{1}{2} \sqrt{\langle \Delta_x^2 \rangle -\langle \Delta_y^2 \rangle }
\end{align}
where the average $\langle ... \rangle$ is over the lattice sites. Notice that this quantity only contributes when there is a nematicity and will vanish for purely extended s-wave and d-wave component of bond or cooper pair density. 

In order to describe our results in the subsequent sections, our strategy is the following. First, we extract the optimal wavevector by comparing the variational ground state energy of the self-consistent BdG solutions at each doping. Subsequently, we utilize the optimal self-consistent solution to study the variation of different order parameters with doping. Finally, we also show the local features of the optimal solutions for a few dopings.
\begin{figure}[h!]
\includegraphics[width=0.475\textwidth]{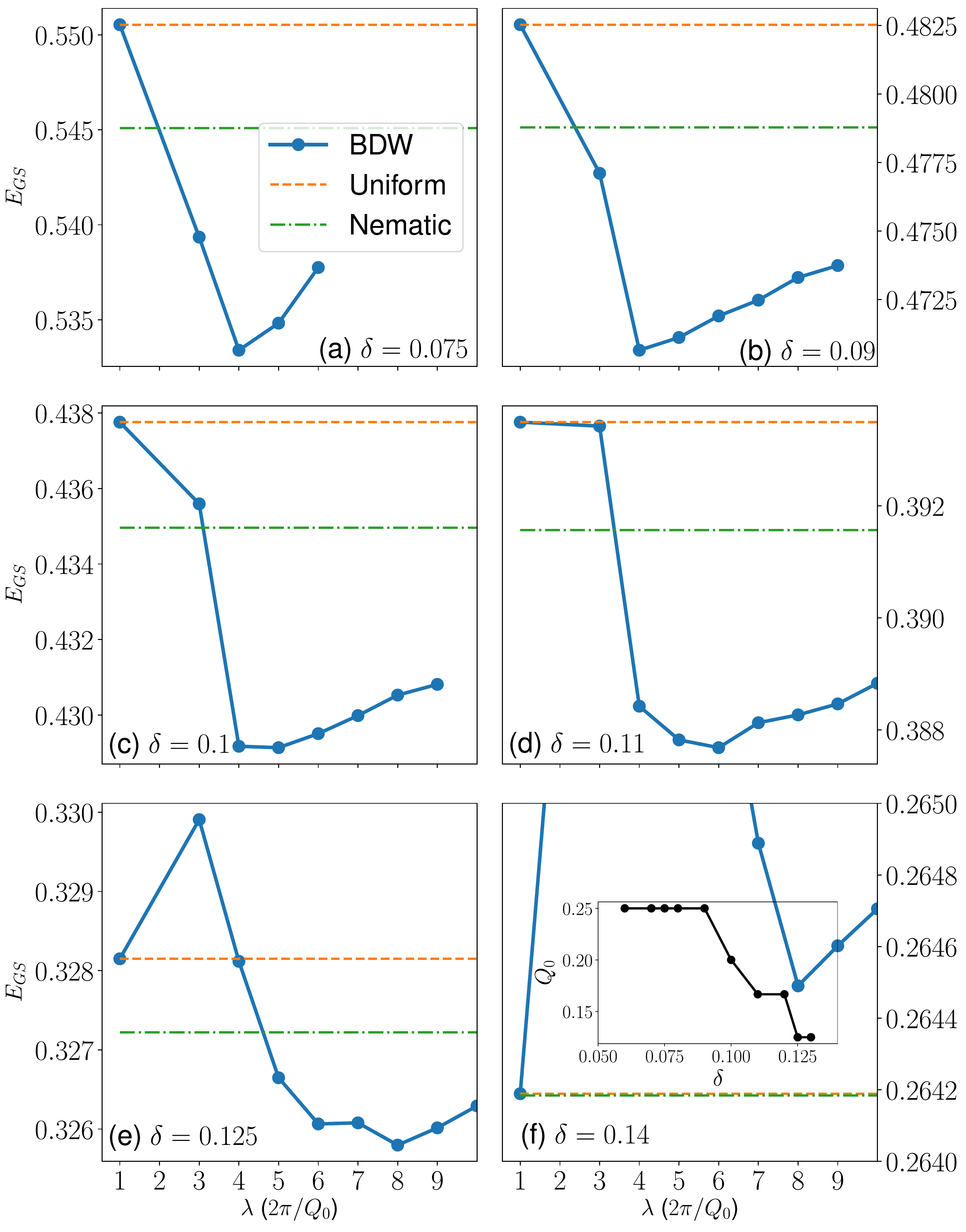}
\caption[0.5\textwidth]{Compares the ground state energy of the self-consistent solutions with different wavevectors for a few dopings. The thick lines are the guide to the eyes. The dotted lines display the $E_{GS}$ for the uniform and the nematic states. (a) For $\delta=0.075$ there is clearly a minima for $\lambda=4$. As the doping rises to $\delta=0.1$, the optimal modulation wavelength increases to $\lambda=5$. The wavelength keeps increasing further with reduced doping, becoming $\lambda=8$ for $\delta=0.125$. At $\delta=0.14$, the nematic state becomes the optimal ground state. The inset of (f) demonstrates the decrease of optimal wavevector with doping.}
\label{fig:fig1}
\end{figure}

\section{Results}
\label{sec:Res}
Our calculations are done on a two-dimensional square lattice unit-cell of linear dimension $40a_0$ to $60a_0$. The lattice size was tuned withing this range to keep the lattice commensurate with the period of modulations. We found that the self-consistency becomes easier if we accommodate even number of full periodic density wave oscillations within the unit-cell. For instance, for a system with period of $8a_0$, we use a system size of $48a_0 \times 48a_0$. We have also checked that our qualitative results are independent of the system size.

\subsection{Optimal modulation wavevector with doping}
\label{sec:EGS}
We begin our analysis by studying the ground state energy for different self-consistent solutions of the BdG equations. Initially, we start with a trail state with a single modulation wavevector at $Q=(1/\lambda) 2\pi/a_0$ for the pairing amplitude, bond and charge density and try to achieve self-consistency. For each case, we achieve self-consistent solutions for the desired wavevector; we calculate the ground state energy ${E_{\rm GS}=\langle \mathcal{H} \rangle_{\rm MF}}$, which is presented in Fig.~(\ref{fig:fig1}) for different wavevectors and for several dopings. The minimum of $E_{\rm GS}(Q)$ gives us the optimal solution at each doping level.

We also present the estimate of the $E_{\rm GS}$ for the uniform and the nematic orders ($Q=0$) with the dashed lines. The nematic order breaks the $\mathcal{C}_4$-rotational symmetry of the square lattice, although it is spatially uniform. Therefore, this is the charge (and pairing) nematic state where the x and the y-component of the bond (and pairing) density are different(See Sec~(\ref{sec:nem}) for details).  
\begin{figure}[h!]
\includegraphics[width=0.475\textwidth]{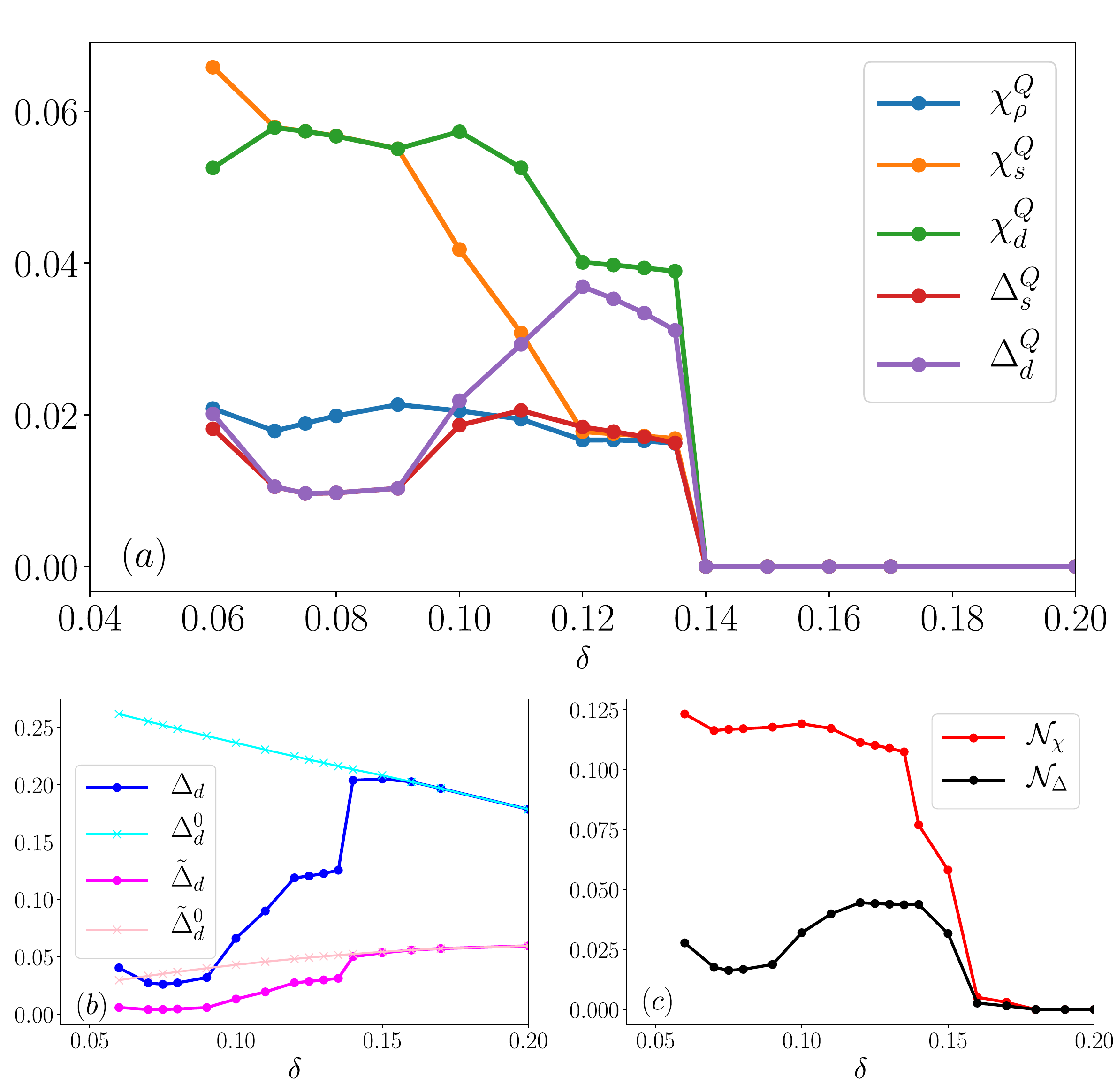}
\caption[0.5\textwidth]{Presents the evolution of different order parameters with doping. (a) Shows the evolution of the amplitude of oscillations for the different quantities at the optimal ground state. The blue lines exhibit the oscillation amplitude for the local density at the optimal wavevector. Similarly, the orange and green curves show the evolution of finite-momentum amplitude for extended $s$-wave and $d$-wave bond density. Moreover, the same for the superconducting pairing in extended $s$-wave and $d$-wave channels is presented in red and violet traces. The lines are the guide to the eyes. All these spatially modulated orders vanish near the $\delta \approx 0.14$. (b) Displays the mean d-wave superconducting pairing in the blue trace, compared with the case where no charge or bond density orders are allowed $\Delta^{0}_{d}$ in cyan. The average $\Delta_d$ significantly reduces in the regions where the charge and bond orders are stable. Moreover, we show the renormalized $d$-wave pairing $\tilde{\Delta}_d$, which also reduces in the regime where charge and pairing oscillations are dominant i.e., $\delta<0.16$. (c) Presents the nematicity in the $\chi$ and $\Delta$ with red and black traces respectively. The nematicity vanishes around the doping $\delta=0.16$.}
\label{fig:fig2}
\end{figure}

At low-doping, the evolution of $E_{\rm GS}$ has a minima at $\lambda=4$, as seen in Fig.~(\ref{fig:fig1}a) and Fig.~(\ref{fig:fig1}b). The purely nematic and the uniform states also have much higher energy than the optimal solution. As the doping is increased to $\delta=0.1$, the energy for a system with density waves of primarily $Q=(1/5) 2\pi/a_0$ minimizes the system. The optimal modulation wavevector shortens further to ${Q_0=1/6-1/8}$ in r.l.u. with increasing doping as seen in Fig.~(\ref{fig:fig1}d) and Fig.~(\ref{fig:fig1}e). The decreasing optimal wavevector with doping is also presented in the inset of Fig.~(\ref{fig:fig1}f). Increasing the doping further makes the nematic state energitically favorable than the spatially modulating ones, as shown in Fig.~(\ref{fig:fig1}f). For $\delta>0.16$, only the uniform SC states are sustained. 

Interestingly, $E_{GS}$ for the doping range $\delta=0.1-0.14$ reveal that states with different wavevectors are quasi-degenerate as marked by the shallowing minima. Such energetically close-lying orders open up the possibility of a state with an admixture of different wavevectors. This can lead to an incommensuration of the density wave~\cite{Vershinin:2004gk,el2018nature}.  We will show in the subsequent sections, that although we start with a single wavevector, higher harmonics of the same wavevector appears in the self-consistent solutions.

Finally, we note that the demise of the modulated orders with doping takes an intriguing pathway. The sharp minima at a particular wavevector at low-dopings indicate that the modulated state with the optimal $Q$ is stable. However, around $\delta=0.14$ as shown in Fig.~(\ref{fig:fig1}f), the modulated states with different wavevectors as well as the nematic and the uniform SC states strikingly become quasi-degenerate . In this regime, minute energy differences can tilt the balance of these orders locally. Beyond $\delta>0.16$, only the uniform $d-$wave superconducting self-consistent solutions are obtained, indicating the demise of modulated and nematic phases.

\subsection{Variation of order parameters with doping}
Above variational analysis identifies the modulated charge patterns in the ground state, but now we proceed to explore the strength of the corresponding spatially modulated orders. In Fig.~(\ref{fig:fig2}a), we show the strength of different modulation amplitude at the dominant wavevector defined using Eq.~(\ref{eq:chi_rho}-\ref{eq:delta_s}). 

Interestingly, over the range of doping, the finite momentum components of BDW, CDW, and PDW orders are significant for $\delta<0.14$ as shown in Fig.~(\ref{fig:fig2}a). However, the d-wave BDW order is the dominant almost over the whole doping range studied. Focusing, on the modulation amplitude for dSC pairing, $\Delta^Q_d$, we observe a non-monotonic behavior with doping. For low doping, the d-wave SC pairing modulations are weaker than other energy scales. However, $\Delta^Q_d$ increase rapidly $\delta=0.1$. Similarly, there remains a significant extended s-wave pairing modulation at all dopings below $\delta=0.14$. We emphasize that the wavevector of the modulated orders reduces around this doping as established in the Sec.~(\ref{sec:EGS}). Recently, STM experiments observed signatures of unidirectional PDW state with $Q=(1/8) 2\pi/a_0$ coexisting with uniform superconducting order in underdoped BSSCO at zero magnetic fields~\cite{Hamidian16,du2020imaging}.

We have also presented the mean d-wave SC pairing amplitude in Fig.~(\ref{fig:fig2}b) and compared it with the situation when all the modulated orders are suppressed in $\Delta^0_d$. In the regime where the spatially varying states exist, the average $\Delta_d$ reduces significantly from the uniform case. The renormalized d-wave pairing amplitude $\tilde{\Delta}_{d}$ also diminishes due to the interplay with the density wave orders.

We also find the signatures of nematicity throughout the doping range $\delta<0.16$, which is indicated by the non-zero $\mathcal{N}_{\Delta}$ and $\mathcal{N}_\chi$ in Fig.~(\ref{fig:fig2}c). The mean bond density and the Cooper pair density is different along the x and y-directions, thus breaking the $\mathcal{C}_4$ rotational symmetry of the square lattice. For the present calculation, the nematic order coexists only with uniform dSC order in the doping range $\delta=0.14-0.16$. However, it also coexists with all spatially modulating orders in the doping range $\delta=0.06-0.14$ as shown in Fig.~(\ref{fig:fig2}). 

\subsection{Spatial profile for the optimal solution}
\label{sec:spatialprofile}
In this subsection, we explore the spatial structure of the optimal solution at a few doping levels.  STM experiments routinely map the spatial variation of diverse orders. Therefore, we start by concentrating on the low-doping regime where we find a unidirectional BDW, PDW, and CDW state with a periodic modulation of four lattice spacing.
\subsubsection{CDW, BDW and PDW state with $Q=(1/4,0)$}
\begin{figure*}[ht]
{\includegraphics[width=16.5cm,keepaspectratio]{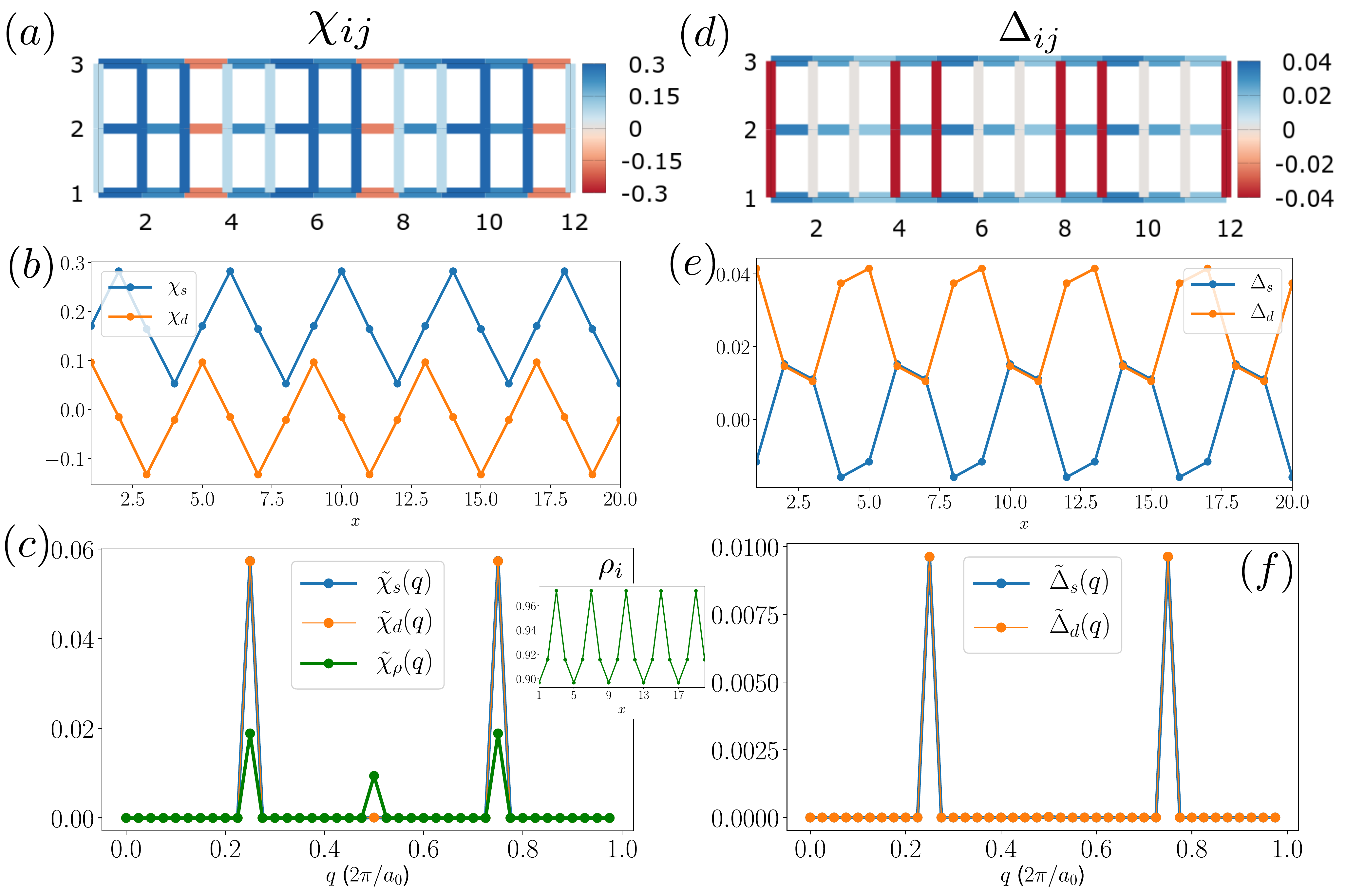}}
\caption{Spatial profile for the optimal self-consistent solution at $\delta=0.075$. Here the periodicity of unidirectional oscillation is $4 a_0$. (a) Shows the bond density $\chi_{ij} =\sum_{\sigma}\langle c^\dagger_{i \sigma} c_{j \sigma} \rangle$, where $i,j$ are nearest neighbouring sites. (b) The d-wave and extended s-wave components of the $\chi$ are extracted in (b). Both clearly show pure oscillations with the periodicity of four lattice spacings. The inset of (c) shows the spatial profile for the local density, which also oscillates. (c) Exhibits the Fourier transform of these quantities. The dominant wavevector for all the oscillations is at $Q=(1/4)2\pi/a_0$ as expected. (d) Shows the similar bond plots for local SC pairing $\Delta_{ij}$. (e) Shows the corresponding $d$-wave and the extended $s$-wave component of SC pairing. (f) Illustrates the Fourier transform for the oscillating part of $\Delta_d$ and $\Delta_s$. The pairing modulations also manifest a dominant peak at $Q=(1/4)2\pi/a_0$ at this dopings. 
}
\label{fig:fig3}
\end{figure*}
The Fig.~(\ref{fig:fig3}a) presents the bond density $\chi_{ij}$ in the color plot. As mentioned before, we consider unidirectional modulated states, and therefore, the bond order only modulates along the x-direction while it is uniform along the y-direction. Interestingly, the $\chi_{ij}$ at each site displays different x and y components as shown in Fig.~(\ref{fig:fig3}a), leading to a charge nematicity coexisting with the density wave orders. 

Next, we extract the $d$-wave and the extended $s$-wave components of the self-consistent Fock shift using Eq.(\ref{eq:chid}) and Eq.~(\ref{eq:chis}). Both the $\chi_d$ and $\chi_s$ modulates in space with a period of four lattice spacings. We also present the variation of the local density $\rho_i$ in the inset of Fig.~($\ref{fig:fig3}$c). We observe spatial modulation of the local density, although it has a much rapid variation instead of regular behavior over a single period.

Fig.~(\ref{fig:fig3}c) exhibits the Fourier transform of bond and charge density as calculated from Eq.~(\ref{eq:chi_rho}-\ref{eq:chi_s}). Our goal is to capture the dominant oscillations for each component of bond and charge modulations. Hence, we removed the mean to eliminate the peak coming from the uniform values of these quantities. The principal peak for the d-wave BDW and s-wave BDW is observed at the expected value of $Q=(1/4) 2\pi/a_0$, and both $\tilde{\chi}_s(Q)$ and $\tilde{\chi}_d(Q)$ have the same strength at this doping. Moreover, there is no other additional peak for both $\tilde{\chi}_d(q)$ and$\tilde{\chi}_s(q)$. The amplitude of the CDW oscillation is much weaker than the BDW ones. Also, there is a sub-dominant peak at $Q=(1/2) 2\pi/a_0$, arising from higher-order harmonics of the local density oscillation.

We present the SC pairing $\Delta_{ij}$ on bonds in Fig.~(\ref{fig:fig3}d) and the extracted d-wave and extended s-wave in Fig.~(\ref{fig:fig3}e). Notice that for a uniform $tJ$ model, the expectation value for the SC pairing with a $d$-wave form-factor is finite, but the same for extended $s$-wave vanishes~\cite{garg2008strong, Chakraborty:2014iq}.  Therefore, the $\Delta_s$ pairing modulation transpires around zero, whereas pairing modulation of the $\Delta_d$ occurs around a non-zero mean value ~\footnote{Note that the presence of the weak nematicity generates a small value of mean $\Delta_s$.}. A PDW state is cooper pairing modulations such that the average pairing vanishes. For the usual pairing modulation due to the interplay of charge or bond density wave order, the mean pairing is significantly larger than the modulation amplitude. Therefore, we observe a PDW oscillation for the extended $s$-wave pairing, whereas a normal pairing modulation in the $d$-wave channel. While the component of extended $s$-wave pairing is essentially absent in the uniform superconductor, the presence of the modulated charge orders brings extended $s$-wave pairing modulation into life. The Fourier transform for the pairing gives an equally strong pairing modulation amplitude for both the form factor with no subdominant peak, as presented in Fig.~(\ref{fig:fig3}d). 

\subsubsection{CDW, BDW and PDW state with $Q=(1/8,0)$}
\begin{figure*}[ht]
{\includegraphics[width=16.5cm,keepaspectratio]{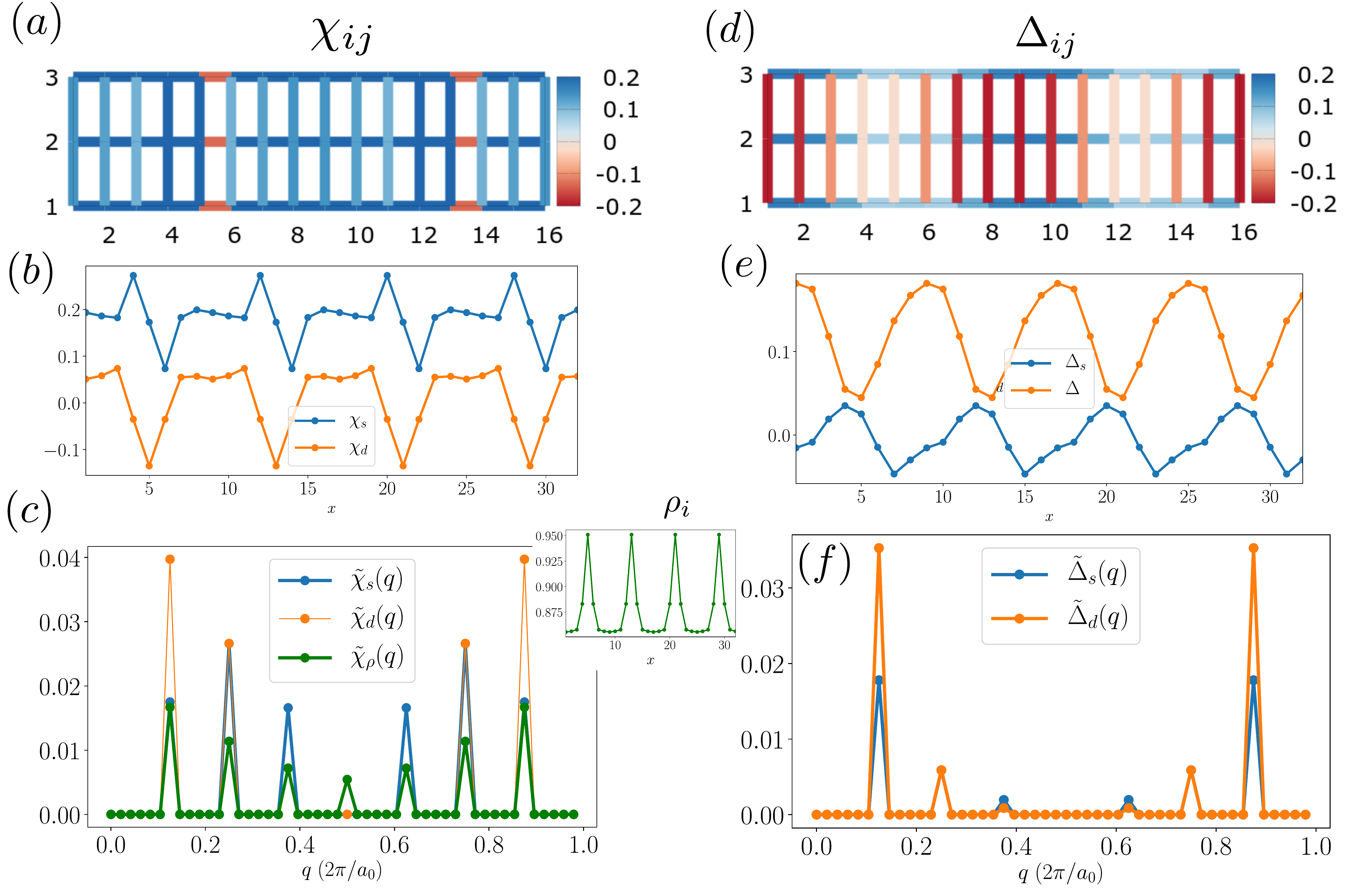}}
\caption{
Spatial profile for the optimal self-consistent solution at $\delta=0.125$. Here the periodicity of unidirectional oscillation is $8 a_0$. (a) Shows the bond density $\chi_{ij}$, where $i,j$ are nearest neighboring sites. (b) The d-wave and extended s-wave components of the $\chi$ are extracted in (b). Both show some jerky oscillations with the periodicity of eight lattice spacings. The inset of (c) shows the spatial profile for the local density, which oscillates. (c) Exhibits the Fourier transform of the quantities mentioned above The dominant wavevector for $\chi_d$ and $\chi_\rho$ are at $Q=(1/8)2\pi/a_0$, whereas the same for $\chi_s$ is at $Q=(1/4)2\pi/a_0$ However, several peaks at the multiples of the ordering wavevector, implies that density modulation is not regular over a period at this doping. (d) Shows the similar bond plots for local SC pairing $\Delta_{ij}$. (e) Shows the corresponding d-wave and the extended s-wave component of SC pairing. (f) Displays the Fourier transform for the oscillating part of $\Delta_d$ and $\Delta_s$. The pairing modulations also exhibit a dominant peak at $Q=(1/8)2\pi/a_0$ at this dopings.
}
\label{fig:fig4}
\end{figure*}
 For $\delta=0.125$, the primary wavevector of the modulated orders in our microscopic model reduces to $Q=(1/8) 2\pi/a_0$. Fig.~(\ref{fig:fig4}a) shows the bond density and  Fig.~(\ref{fig:fig4}b) the $d$-wave and $s^\prime$-wave component of the same. The modulation of the bond density is not smooth over the period and therefore accommodates higher harmonics. This is also true for the density modulations presented as an inset of Fig.~(\ref{fig:fig4}c). The presence of higher harmonics of oscillations is verified by the strong subdominant peaks of the Fourier transform of the BDW and CDW orders as exhibited in Fig.~(\ref{fig:fig4}c). Interestingly, the $d$-wave BDW becomes the dominant form factor at $q=0.125$ in reciprocal lattice unit (rlu). However, the $\tilde{\chi}_s$ has an intriguing evolution. For the extended $s$-wave BDW modulations, the dominant peak shifts at $q=(1/4) 2\pi/a_0$, which is at $\mathbf{q}=2\mathbf{Q}$. 
 
 Similarly, the SC pairing amplitude is presented in Fig.~(\ref{fig:fig4}d), and the extended $s$-wave and $d$-wave components are presented in Fig.~(\ref{fig:fig4}e). The pairing modulation is smoother contrasted to the BDW or CDW oscillations. In Fig.~(\ref{fig:fig4}f) the Fourier transform of $\Delta_d$ and $\Delta_s$ confirms the dominant ordering at $q=(1/8) 2\pi/a_0$. The secondary peaks are also much weaker than the principal ones.
 
 \subsubsection{Nematic state}
 \label{sec:nem}
\begin{figure}[h!]
\includegraphics[width=0.475\textwidth]{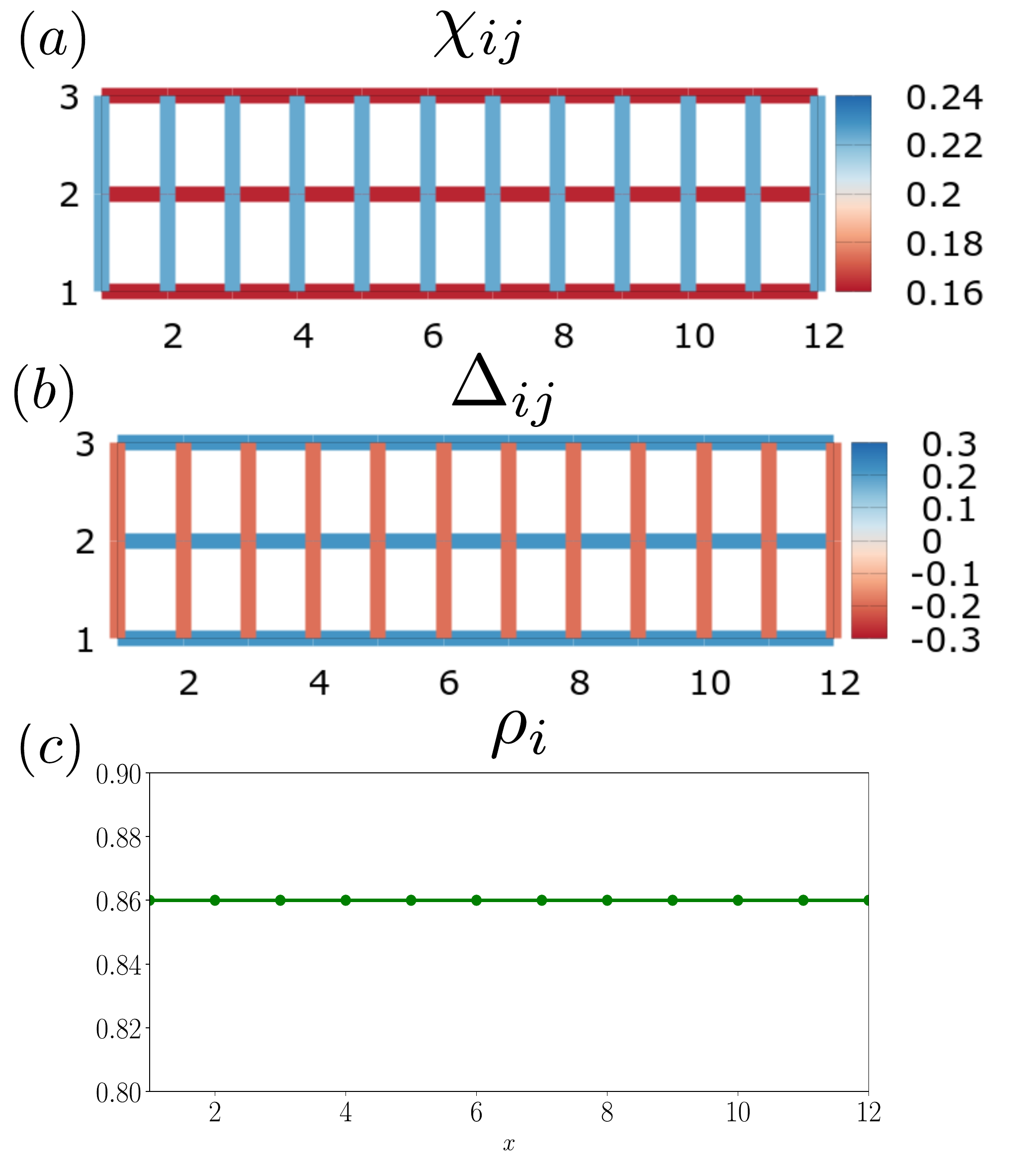}
\caption[0.5\textwidth]{Shows the spatial dependence for the optimal self-consistent solution at $\delta=0.14$. Although the optimal self-consistent solution has no modulation in this parameter regime, it breaks the $\mathcal{C}_4$-rotational symmetry of the square lattice leading to a charge nematic state. (a) Shows the bond density $\chi_{ij}$, which is different for the x-bonds and y-bonds. Similarly, the local SC pairing is not purely d-wave as shown in (b) as expected in the uniform tJ-model. Interestingly, there is no spatial variation of these quantities or the local density as presented in (c).}
\label{fig:fig5}
\end{figure}
Finally, for $\delta=0.14$, the optimal self-consistent solution supports a $\mathbf{Q}=0$ nematic state. Fig.~(\ref{fig:fig5}a) shows the bond density amplitude. The color difference indicates the breaking of the rotational symmetry of the square lattice. However, there are no modulations of the bond density either along the x or y-directions. Moreover, the lack of spatially modulated order is confirmed by the SC pairing in Fig.~(\ref{fig:fig5}b) and the local density in Fig.~(\ref{fig:fig5}c). 

\section{Discussions}
\label{sec:disc}
We study the $t-t^{\prime}-J$ model with additional nearest-neighbor repulsion and find modulating bond density wave states coexisting with charge and pairing modulated states in the underdoped regime. The modulated states have lower energy than the uniform $d$-wave superconducting state in the doping range $\delta=0.06-0.14$. The optimal wavevector for all the modulated quantities is commensurate with the lattice with higher harmonics of oscillations. Over the doping range $\delta=0.06-0.16$, there remains a significant nematicity such that the $\mathcal{C}_4$-rotational symmetry is broken. Moreover, for $\delta=0.14-0.16$, the system supports a purely nematic state without any translation symmetry breaking. Finally, beyond this doping, only a uniform $d-$wave superconducting order is stabilized. In the following subsections, we discuss how this work compares with other results in the literature in the context of the cuprates.

\subsection{Role of strong correlation in stabilizing modulated orders}
We attempted to study the modulated states by relaxing the strong correlation constraint by setting the Gutzwiller factors to unity. In the past, studies have been performed for d-wave SC in the $tJ$ model, and this procedure is known as Inhomogeneous Mean Field Theory (IMT)~\cite{garg2008strong,Chakraborty:2014iq}. In sharp contrast to the strongly correlated results presented in this paper, we failed to stabilize any modulated solutions across the parameter regime studied. Of course, an absence of modulated self-consistent solutions might be due to a lack of exhaustive search or a bad initial guess of the order parameters. However, any initial trail state with modulated charge, pairing, and bond densities flowed strictly towards a uniform $d$-wave SC solution. Furthermore, the ease of achieving modulated self-consistent solutions for multiple wavevectors over a wide doping range in the highly correlated regime indicates a lack of modulated states in the absence of a strong correlation. The propensity to form modulated orders near the half-filling where the interactions are stronger also hints at the same possibility. Recent explorations of the pair density of states in the $tJ-$ model also observed a lack of modulated orders when the Gutzwiller factors are set to unity~\cite{tu2019}.

\subsection{Evolution of optimal wavevector with doping}
In the $t-t^{\prime}-J$ model with additional nearest-neighbor repulsion, the optimal modulation wavevector reduces with increasing doping. For doping around $\delta=0.06-0.09$ the optimal wavevector is around $Q=(1/4) 2\pi/a_0$. As the doping is increased further, the periodicity increases to $8a_0$ around $\delta=0.125$ as shown in the inset of Fig.~(\ref{fig:fig1}f). Generally, in the cuprates, the wavevectors also diminish with increased hole doping~\cite{Comin14,daSilvaNeto14} except in La-based material~\cite{Fink09,Fink11,Hucker11}. However, in \chem{La_2CuO_4} the additional spin density wave play a vital role in dictating the modulation wavevector~\cite{VestigeNie,miao17}. It would be interesting to discern the role of magnetic orders on the modulation wavevector in the future. 

Additionally, in cuprates the wavevector varies from ${Q=0.33-0.2}$ $(2\pi/a_0)$ leading to a periodicity of $3-5$ lattice spacing up to $\delta=0.2$. However, this contrasts with theoretical calculations that often predict longer modulation wavelength in the $tJ$-model and the associated Hubbard model at large $U$~\cite{HubbardStripe17}. The evolution of the ground state energy in Fig.~(\ref{fig:fig1}) hints towards an explanation of this apparent anomaly. Notice that below $\delta<0.1$, the difference between the $E_{GS}$ at the optimal wavevector with other wavevector is significant. Note that we only studied the commensurate wavelengths with the lattice spacings. However, the energy difference between the states with different wavelengths becomes minute, $\delta E \sim 0.001t$ as doping increases beyond $\delta>0.1$. Since these states are near degenerate, various fluctuations absent in our study can modify the delicate balance of energy. Consequently, the state with a smaller wavelength can become the ground state. 

Aditionally, there can be another important consequnece for near-degeneracy of modulated states with different wavevectors as shown in Fig.~(\ref{fig:fig1}c)-Fig.~(\ref{fig:fig1}f). This can possibly support a state with an adimixture of different wavevectors. Furthermore, the irregular evolution of the $\chi_d$ and $\chi_s$ over a period in Fig.~(\ref{fig:fig4}b) also hints towards the incipient incommensuration. Such patterns generates strong higher harmonics in the Fourier transform of the BDW order in Fig.~(\ref{fig:fig4}c). A splitting or broadening of the primary peak of the $\tilde{\chi}(q)$ indicates an incommensuration of any density wave pattern. Here, strong higher harmonics with the presence of nearby quasi-degenerate states can drive a multi-peak incommensuration around  $\delta \approx 0.125$. 

\subsection{Nearest neigbor repulsion term in the $tJ$-model}
The $t-J$ model and its ``parent" -- Hubbard model with large on-site repulsion $U$ are generally considered the minimal model for cuprate superconductors. However, the interactions are not entirely screened in realistic materials to be completely local. A residual nearest-neighbor repulsion is estimated to be one order of magnitude smaller
than the on-site repulsion in cuprates~\cite{VtermImp0,VtermImp1,VtermImp2}. Such interactions are often neglected due to their reduced magnitude. However, they become crucial when the ground state without them shows near-degeneracy among several broken symmetry states, as in the underdoped regime of the $t-J$ model.

The modulated orders in the $t-t^\prime-J$ model have been studied extensively in the past~\cite{yang2009nature,choubey2017,tu2019,ZegrodnikPRB}. These studies found different modulating bond, charge, and pair density wave states energetically very close yet higher than the uniform superconducting ground state. Therefore, additional ingredients absent in the $t-t^\prime-J$
must be responsible for stabilizing the spatially modulating density wave orders as the ground state in cuprates. 
Ref.~\cite{ZegrodnikPRB} suggests studying a $tJ$-model with additional on-site repulsion stabilizing nematic phases with superconductivity.
There are other studies suggesting that adding a nearest-neighbor repulsion term to prefer the density wave orders as the ground state~\cite{Allais14c,spalek2017}.
 In particular, the nearest neighbor repulsion is expected to be detrimental to the $d$-wave superconductivity and can thus stabilize the nearby density waves orders as the ground state. Similar to these studies, we also observed that a critical value of nearest neighbor repulsion is essential to stabilize the density wave orders~\cite{Allais14c,spalek2017}.
 
Furthermore, increasing the nearest neighbor repulsion term keeping the other terms constant allows for the modulated orders to survive up to higher dopings. We have confirmed that for $W=0.75t$, the spatially modulated orders remain the ground state up to $\delta=0.16$ whereas the ground state becomes purely nematic in $\delta=0.16-0.19$. The energy difference between nematic and the uniform state remains significantly more prominent for larger $W$ leading to unambiguous identification of the nematic regime.

\subsection{Pairing modulation}
We observe modulations of the cooper pair density both in the extended $s$-wave and the $d$-wave channel over the doping range $\delta=0.06-0.14$. A modulation of cooper pairing can be a direct consequence of the coexistence with the BDW order. On one hand, the modulation of the bond density leads to an oscillation of the local density of states which dictates the propensity for developing local SC pair and drives the modulation of the pairing. On the other hand, a more exotic pair density wave is a state of spatially oscillating Cooper pair such that the average pairing vanishes or is very small. In our analysis, the $\Delta_s$ shows a vanishingly small mean with a robust modulation amplitude over the whole doping range as presented in Fig.~(\ref{fig:fig3}e) and Fig.~(\ref{fig:fig4}e). On the other hand, the $d$-wave SC pairing modulates around a finite non-zero value. Consequently, this indicates a PDW with $s^\prime$ form factor coexisting with other modulated states over a wide doping range.

Recent experiments on BSSCO~\cite{Hamidian16,du2020imaging} claims to observe PDW modulations at $\mathbf{Q}_P=(1/8) 2\pi/a_0$ of $\sim 6$ meV coexisting with a uniform SC at $\sim 36$ meV. The observed PDW state has either extended $s$-wave or $s$-wave form factor. These studies also observe CDW oscillations at $\mathbf{Q}_C=\mathbf{Q}_p$ and at $\mathbf{Q}_C=2\mathbf{Q}_P$.  The model calculations around $\delta=0.125$ reveal a similar behavior -- an extended $s$-wave PDW undulation with a non-zero mean $d$-wave superconductor. Furthermore, the optimal self-consistent solution displays a PDW order $\mathbf{q}=\mathbf{Q}$ with the CDW orders at both $\mathbf{q}=\mathbf{Q}$ and $\mathbf{q}=2\mathbf{Q}$, very similar to what we observe in Fig.~(\ref{fig:fig4}c). It is encouraging that our simplified model for modulated orders in strongly correlated systems can capture some features of the experimental signatures.
\begin{figure}[h!]
\includegraphics[width=0.475\textwidth]{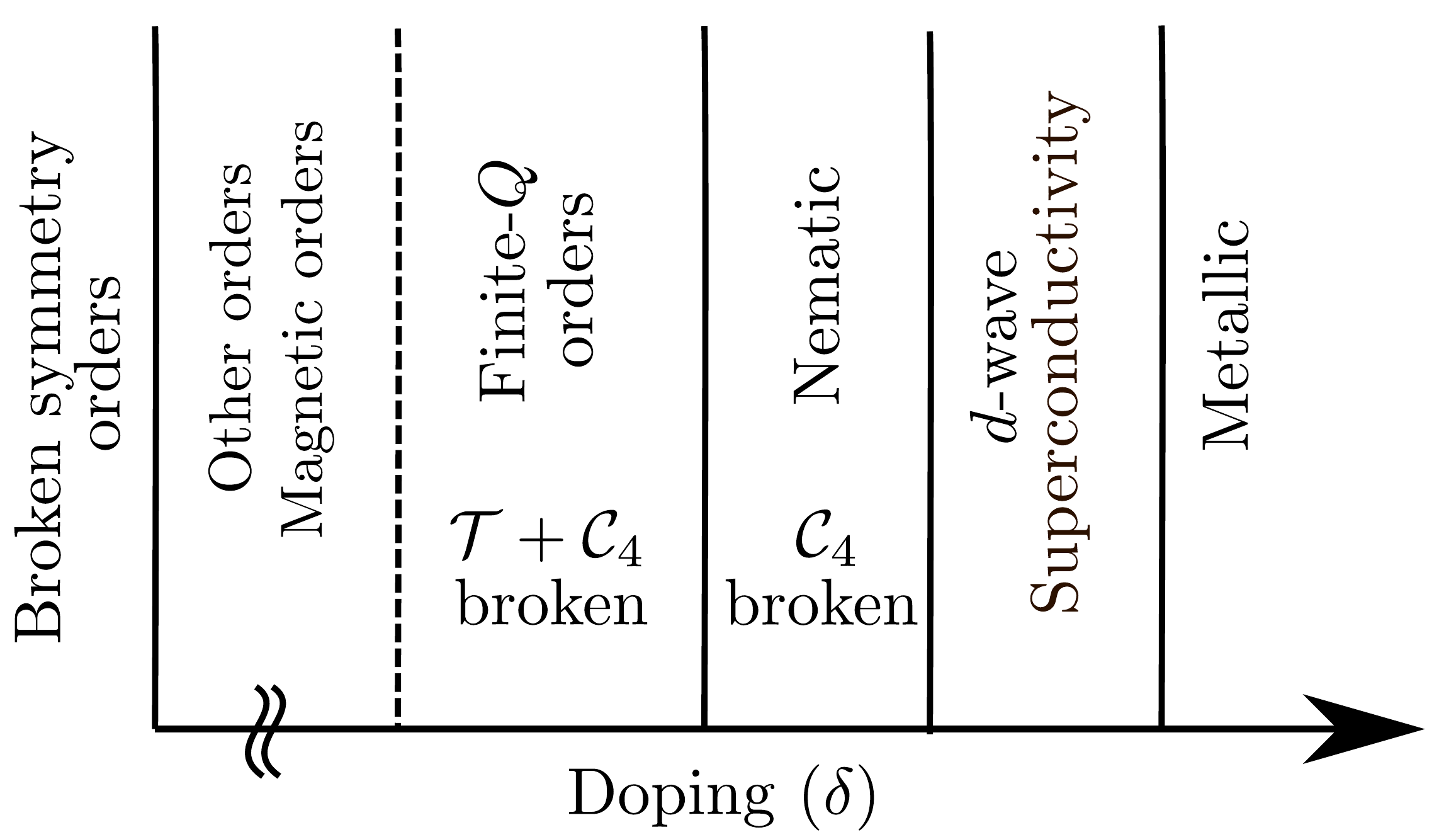}
\caption[0.5\textwidth]{Presents a schematic symmetry broken phases with doping from our model calculations at zero temperature. The doping regime presented below the broken line is beyond the scope of this study. Below this doping we expect magenetic and other symmetries to be broken. From the low to intermediate dopings, both the translation symmetry, $\mathcal{T}$ and the $\mathcal{C}_4$ rotational symmetry is broken, leading to the observation of the bond, charge, and pair density wave orders. Beyond a doping level, only $\mathcal{C}_4$ rotational symmetry is broken, leading to a nematic state. Note that fewer symmetries are broken as the doping is increased. Beyond critical doping, only translational and rotational invariant d-wave SC remains robust. Uniform superconductors only break the particle number conservation symmetry. The uniform $d$-wave SC remains robust over the whole doping regime studied, only decaying at much higher doping.}
\label{fig:fig6}
\end{figure}
\subsection{Vestigial nematic order}
Our model calculation reveals that a translationally invariant nematic state stabilizes in the doping range $\delta=0.14-0.16$. Note that the nematic state is a higher order component of the superconducting and bond order parameter. If the bond order $\chi$ is written as a vector then the nematic order is given by $N_{\chi}=1/2(\langle \chi^\dagger \sigma_z \chi \rangle$, where $\sigma_z$ is the z-component of the Pauli matrix. In the nematic state the combined order parameter is finite although the $\langle \chi^Q_x \rangle=0$ and $\langle \chi^Q_y \rangle=0$. Consequently, nematic state breaks only a subset of the all possible symmetry that can be broken in this regime leading to a vestigial nematic state.

Recently sublattice resolved electronic structure experiments in cuprates~\cite{nematic19PNAS} reveal a $Q=0$ nematic state over the doping range $\delta=0.06-0.17$. The vestigial nematic state coexists with finite-$Q$ $d$-wave charge order and a uniform $d$-wave in the intermediate dopings. A pure nematic state without any translational symmetry breaking is also found in the doping range $\delta=0.15-0.17$. We note that our optimal solution also sustains a pure $\mathcal{C}_4$ rotational symmetry breaking order around doping. Beyond this doping, only uniform d-wave superconductivity stabilizes in our model calculation. 

\subsection{Phase diagram}
In Fig.~(\ref{fig:fig6}), we provide a schematic map of different symmetry breaking orders based on our model calculations at zero temperature. The doping regime presented below the broken line is beyond the scope of this study. Uniform superconductors only break the particle number conservation symmetry. At zero temperature, our model calculation suggests that uniform $d$-wave SC remains robust over the whole doping regime studied, only decaying at much higher doping. In contrast, the finite-$Q$ density wave states become relevant only in the lightly doped regime. The density wave order breaks both the translation, $\mathcal{T}$ and the $\mathcal{C}_4$ rotational symmetry of the square lattice. Note that Fig.~(\ref{fig:fig2}b) shows that the mean $d$-wave pairing amplitude reduces drastically in the regime where other symmetries are broken.  Consequently, the suppression of SC pairing is expected from our calculations at $T=0$. A similar drop in superconducting correlations is observed in cuprates around the hole-doping of $12\%$, where the amplitude of the modulated orders is strong~\cite{Huecker14a,Blanco-Canosa14}.

Between the finite-$Q$ orders and the uniform SC, there is a regime where the rotational symmetry remains broken although the translational invariance is restored. Such a state is a vestigial nematic state which breaks a a smaller subset of symmetries than the primary state of density wave orders. The series of cascading broken symmetries as the doping reduces hints at the possible entanglement between multiple orders near the highly correlated regime.

The charge density waves~\cite{Loret19} and the associated vestigial nematicity~\cite{nematic19PNAS, Nie13, Vestigial19_ARCMP} are considered an essential ingredient of the pseudogap state. Our study suggests that ground state of cuprates in the underdoped regime indeed breaks certain symmetries generating various density wave states, which manifests in experiments as the pseudogap state in cuprates at finite temperature.
Therefore, to capture the pseudogap state, it is essential to analyze the effects of different fluctuations on all different broken symmetry states.
Furthermore, thermal and quantum fluctuations can lead to spatial reorganization into domains of finite-$Q$ and $Q=0$ orders~\cite{Lee21,Banerjee18,ag08dislocations}. Additionally, the Mott insulator and magnetically ordered states can significantly alter the states at low doping. A rigorous treatment of the magnetic orders with these broken symmetry states, along with the effects of thermal fluctuations, remains to be explored in the future.

\section{Acknowledgement}
The authors thank Yvan Sidis, Maxence Grandadam, and Debmalya Chakraborty for valuable discussions. This work has received financial support from the ERC, under grant agreement AdG694651-CHAMPAGNE.
\bibliography{Cuprates.bib}

\begin{thebibliography}{70}%
\makeatletter
\providecommand \@ifxundefined [1]{%
 \@ifx{#1\undefined}
}%
\providecommand \@ifnum [1]{%
 \ifnum #1\expandafter \@firstoftwo
 \else \expandafter \@secondoftwo
 \fi
}%
\providecommand \@ifx [1]{%
 \ifx #1\expandafter \@firstoftwo
 \else \expandafter \@secondoftwo
 \fi
}%
\providecommand \natexlab [1]{#1}%
\providecommand \enquote  [1]{``#1''}%
\providecommand \bibnamefont  [1]{#1}%
\providecommand \bibfnamefont [1]{#1}%
\providecommand \citenamefont [1]{#1}%
\providecommand \href@noop [0]{\@secondoftwo}%
\providecommand \href [0]{\begingroup \@sanitize@url \@href}%
\providecommand \@href[1]{\@@startlink{#1}\@@href}%
\providecommand \@@href[1]{\endgroup#1\@@endlink}%
\providecommand \@sanitize@url [0]{\catcode `\\12\catcode `\$12\catcode
  `\&12\catcode `\#12\catcode `\^12\catcode `\_12\catcode `\%12\relax}%
\providecommand \@@startlink[1]{}%
\providecommand \@@endlink[0]{}%
\providecommand \url  [0]{\begingroup\@sanitize@url \@url }%
\providecommand \@url [1]{\endgroup\@href {#1}{\urlprefix }}%
\providecommand \urlprefix  [0]{URL }%
\providecommand \Eprint [0]{\href }%
\providecommand \doibase [0]{http://dx.doi.org/}%
\providecommand \selectlanguage [0]{\@gobble}%
\providecommand \bibinfo  [0]{\@secondoftwo}%
\providecommand \bibfield  [0]{\@secondoftwo}%
\providecommand \translation [1]{[#1]}%
\providecommand \BibitemOpen [0]{}%
\providecommand \bibitemStop [0]{}%
\providecommand \bibitemNoStop [0]{.\EOS\space}%
\providecommand \EOS [0]{\spacefactor3000\relax}%
\providecommand \BibitemShut  [1]{\csname bibitem#1\endcsname}%
\let\auto@bib@innerbib\@empty
\bibitem [{\citenamefont {Keimer}\ and\ \citenamefont
  {Moore}(2017)}]{keimer2017}%
  \BibitemOpen
  \bibfield  {author} {\bibinfo {author} {\bibfnamefont {B.}~\bibnamefont
  {Keimer}}\ and\ \bibinfo {author} {\bibfnamefont {J.}~\bibnamefont {Moore}},\
  }\href@noop {} {\bibfield  {journal} {\bibinfo  {journal} {Nature Physics}\
  }\textbf {\bibinfo {volume} {13}},\ \bibinfo {pages} {1045} (\bibinfo {year}
  {2017})}\BibitemShut {NoStop}%
\bibitem [{\citenamefont {Sun}\ \emph {et~al.}(2016)\citenamefont {Sun},
  \citenamefont {Matsuura}, \citenamefont {Ye}, \citenamefont {Mizukami},
  \citenamefont {Shimozawa}, \citenamefont {Matsubayashi}, \citenamefont
  {Yamashita}, \citenamefont {Watashige}, \citenamefont {Kasahara},
  \citenamefont {Matsuda}, \citenamefont {Yan}, \citenamefont {Sales},
  \citenamefont {Uwatoko}, \citenamefont {Cheng},\ and\ \citenamefont
  {Shibauchi}}]{sun2016dome}%
  \BibitemOpen
  \bibfield  {author} {\bibinfo {author} {\bibfnamefont {J.~P.}\ \bibnamefont
  {Sun}}, \bibinfo {author} {\bibfnamefont {K.}~\bibnamefont {Matsuura}},
  \bibinfo {author} {\bibfnamefont {G.~Z.}\ \bibnamefont {Ye}}, \bibinfo
  {author} {\bibfnamefont {Y.}~\bibnamefont {Mizukami}}, \bibinfo {author}
  {\bibfnamefont {M.}~\bibnamefont {Shimozawa}}, \bibinfo {author}
  {\bibfnamefont {K.}~\bibnamefont {Matsubayashi}}, \bibinfo {author}
  {\bibfnamefont {M.}~\bibnamefont {Yamashita}}, \bibinfo {author}
  {\bibfnamefont {T.}~\bibnamefont {Watashige}}, \bibinfo {author}
  {\bibfnamefont {S.}~\bibnamefont {Kasahara}}, \bibinfo {author}
  {\bibfnamefont {Y.}~\bibnamefont {Matsuda}}, \bibinfo {author} {\bibfnamefont
  {J.-Q.}\ \bibnamefont {Yan}}, \bibinfo {author} {\bibfnamefont {B.~C.}\
  \bibnamefont {Sales}}, \bibinfo {author} {\bibfnamefont {Y.}~\bibnamefont
  {Uwatoko}}, \bibinfo {author} {\bibfnamefont {J.-G.}\ \bibnamefont {Cheng}},
  \ and\ \bibinfo {author} {\bibfnamefont {T.}~\bibnamefont {Shibauchi}},\
  }\href {\doibase 10.1038/ncomms12146} {\bibfield  {journal} {\bibinfo
  {journal} {Nature Communications}\ }\textbf {\bibinfo {volume} {7}},\
  \bibinfo {pages} {12146} (\bibinfo {year} {2016})}\BibitemShut {NoStop}%
\bibitem [{\citenamefont {Dagotto}(2005)}]{dagotto2005complexity}%
  \BibitemOpen
  \bibfield  {author} {\bibinfo {author} {\bibfnamefont {E.}~\bibnamefont
  {Dagotto}},\ }\href@noop {} {\bibfield  {journal} {\bibinfo  {journal}
  {Science}\ }\textbf {\bibinfo {volume} {309}},\ \bibinfo {pages} {257}
  (\bibinfo {year} {2005})}\BibitemShut {NoStop}%
\bibitem [{\citenamefont {Bounoua}\ \emph {et~al.}(2020)\citenamefont
  {Bounoua}, \citenamefont {Mangin-Thro}, \citenamefont {Jeong}, \citenamefont
  {Saint-Martin}, \citenamefont {Pinsard-Gaudart}, \citenamefont {Sidis},\ and\
  \citenamefont {Bourges}}]{bounoua2020loop}%
  \BibitemOpen
  \bibfield  {author} {\bibinfo {author} {\bibfnamefont {D.}~\bibnamefont
  {Bounoua}}, \bibinfo {author} {\bibfnamefont {L.}~\bibnamefont
  {Mangin-Thro}}, \bibinfo {author} {\bibfnamefont {J.}~\bibnamefont {Jeong}},
  \bibinfo {author} {\bibfnamefont {R.}~\bibnamefont {Saint-Martin}}, \bibinfo
  {author} {\bibfnamefont {L.}~\bibnamefont {Pinsard-Gaudart}}, \bibinfo
  {author} {\bibfnamefont {Y.}~\bibnamefont {Sidis}}, \ and\ \bibinfo {author}
  {\bibfnamefont {P.}~\bibnamefont {Bourges}},\ }\href@noop {} {\bibfield
  {journal} {\bibinfo  {journal} {Communications Physics}\ }\textbf {\bibinfo
  {volume} {3}},\ \bibinfo {pages} {1} (\bibinfo {year} {2020})}\BibitemShut
  {NoStop}%
\bibitem [{\citenamefont {Fradkin}\ \emph {et~al.}(2015)\citenamefont
  {Fradkin}, \citenamefont {Kivelson},\ and\ \citenamefont
  {Tranquada}}]{Fradkin:2015ch}%
  \BibitemOpen
  \bibfield  {author} {\bibinfo {author} {\bibfnamefont {E.}~\bibnamefont
  {Fradkin}}, \bibinfo {author} {\bibfnamefont {S.~A.}\ \bibnamefont
  {Kivelson}}, \ and\ \bibinfo {author} {\bibfnamefont {J.~M.}\ \bibnamefont
  {Tranquada}},\ }\href {\doibase 10.1103/revmodphys.87.457} {\bibfield
  {journal} {\bibinfo  {journal} {Rev. Mod. Phys.}\ }\textbf {\bibinfo {volume}
  {87}},\ \bibinfo {pages} {457} (\bibinfo {year} {2015})}\BibitemShut
  {NoStop}%
\bibitem [{\citenamefont {Fernandes}\ \emph {et~al.}(2019)\citenamefont
  {Fernandes}, \citenamefont {Orth},\ and\ \citenamefont
  {Schmalian}}]{Vestigial19_ARCMP}%
  \BibitemOpen
  \bibfield  {author} {\bibinfo {author} {\bibfnamefont {R.~M.}\ \bibnamefont
  {Fernandes}}, \bibinfo {author} {\bibfnamefont {P.~P.}\ \bibnamefont {Orth}},
  \ and\ \bibinfo {author} {\bibfnamefont {J.}~\bibnamefont {Schmalian}},\
  }\href@noop {} {\bibfield  {journal} {\bibinfo  {journal} {Annual Review of
  Condensed Matter Physics}\ }\textbf {\bibinfo {volume} {10}},\ \bibinfo
  {pages} {133} (\bibinfo {year} {2019})}\BibitemShut {NoStop}%
\bibitem [{\citenamefont {Nie}\ \emph {et~al.}(2014)\citenamefont {Nie},
  \citenamefont {Tarjus},\ and\ \citenamefont {Kivelson}}]{Nie13}%
  \BibitemOpen
  \bibfield  {author} {\bibinfo {author} {\bibfnamefont {L.}~\bibnamefont
  {Nie}}, \bibinfo {author} {\bibfnamefont {G.}~\bibnamefont {Tarjus}}, \ and\
  \bibinfo {author} {\bibfnamefont {S.~A.}\ \bibnamefont {Kivelson}},\ }\href
  {\doibase 10.1073/pnas.1406019111} {\bibfield  {journal} {\bibinfo  {journal}
  {Proc. Natl. Acad. Sci.}\ }\textbf {\bibinfo {volume} {111}},\ \bibinfo
  {pages} {7980} (\bibinfo {year} {2014})}\BibitemShut {NoStop}%
\bibitem [{\citenamefont {Mukhopadhyay}\ \emph {et~al.}(2019)\citenamefont
  {Mukhopadhyay}, \citenamefont {Sharma}, \citenamefont {Kim}, \citenamefont
  {Edkins}, \citenamefont {Hamidian}, \citenamefont {Eisaki}, \citenamefont
  {Uchida}, \citenamefont {Kim}, \citenamefont {Lawler}, \citenamefont
  {Mackenzie} \emph {et~al.}}]{nematic19PNAS}%
  \BibitemOpen
  \bibfield  {author} {\bibinfo {author} {\bibfnamefont {S.}~\bibnamefont
  {Mukhopadhyay}}, \bibinfo {author} {\bibfnamefont {R.}~\bibnamefont
  {Sharma}}, \bibinfo {author} {\bibfnamefont {C.~K.}\ \bibnamefont {Kim}},
  \bibinfo {author} {\bibfnamefont {S.~D.}\ \bibnamefont {Edkins}}, \bibinfo
  {author} {\bibfnamefont {M.~H.}\ \bibnamefont {Hamidian}}, \bibinfo {author}
  {\bibfnamefont {H.}~\bibnamefont {Eisaki}}, \bibinfo {author} {\bibfnamefont
  {S.-i.}\ \bibnamefont {Uchida}}, \bibinfo {author} {\bibfnamefont {E.-A.}\
  \bibnamefont {Kim}}, \bibinfo {author} {\bibfnamefont {M.~J.}\ \bibnamefont
  {Lawler}}, \bibinfo {author} {\bibfnamefont {A.~P.}\ \bibnamefont
  {Mackenzie}},  \emph {et~al.},\ }\href@noop {} {\bibfield  {journal}
  {\bibinfo  {journal} {Proceedings of the National Academy of Sciences}\
  }\textbf {\bibinfo {volume} {116}},\ \bibinfo {pages} {13249} (\bibinfo
  {year} {2019})}\BibitemShut {NoStop}%
\bibitem [{\citenamefont {Hoffman}\ \emph {et~al.}(2002)\citenamefont
  {Hoffman}, \citenamefont {Hudson}, \citenamefont {Lang}, \citenamefont
  {Madhavan}, \citenamefont {Eisaki}, \citenamefont {Uchida},\ and\
  \citenamefont {Davis}}]{Hoffman02}%
  \BibitemOpen
  \bibfield  {author} {\bibinfo {author} {\bibfnamefont {J.~E.}\ \bibnamefont
  {Hoffman}}, \bibinfo {author} {\bibfnamefont {E.~W.}\ \bibnamefont {Hudson}},
  \bibinfo {author} {\bibfnamefont {K.~M.}\ \bibnamefont {Lang}}, \bibinfo
  {author} {\bibfnamefont {V.}~\bibnamefont {Madhavan}}, \bibinfo {author}
  {\bibfnamefont {H.}~\bibnamefont {Eisaki}}, \bibinfo {author} {\bibfnamefont
  {S.}~\bibnamefont {Uchida}}, \ and\ \bibinfo {author} {\bibfnamefont {J.~C.}\
  \bibnamefont {Davis}},\ }\href {\doibase 10.1126/science.1066974} {\bibfield
  {journal} {\bibinfo  {journal} {Science}\ }\textbf {\bibinfo {volume}
  {295}},\ \bibinfo {pages} {466} (\bibinfo {year} {2002})}\BibitemShut
  {NoStop}%
\bibitem [{\citenamefont {Wise}\ \emph {et~al.}(2008)\citenamefont {Wise},
  \citenamefont {Boyer}, \citenamefont {Chatterjee}, \citenamefont {Kondo},
  \citenamefont {Takeuchi}, \citenamefont {Ikuta}, \citenamefont {Wang},\ and\
  \citenamefont {Hudson}}]{Wise08}%
  \BibitemOpen
  \bibfield  {author} {\bibinfo {author} {\bibfnamefont {W.~D.}\ \bibnamefont
  {Wise}}, \bibinfo {author} {\bibfnamefont {M.~C.}\ \bibnamefont {Boyer}},
  \bibinfo {author} {\bibfnamefont {K.}~\bibnamefont {Chatterjee}}, \bibinfo
  {author} {\bibfnamefont {T.}~\bibnamefont {Kondo}}, \bibinfo {author}
  {\bibfnamefont {T.}~\bibnamefont {Takeuchi}}, \bibinfo {author}
  {\bibfnamefont {H.}~\bibnamefont {Ikuta}}, \bibinfo {author} {\bibfnamefont
  {Y.}~\bibnamefont {Wang}}, \ and\ \bibinfo {author} {\bibfnamefont {E.~W.}\
  \bibnamefont {Hudson}},\ }\href {http://dx.doi.org/10.1038/nphys1021}
  {\bibfield  {journal} {\bibinfo  {journal} {Nat. Phys.}\ }\textbf {\bibinfo
  {volume} {4}},\ \bibinfo {pages} {696} (\bibinfo {year} {2008})}\BibitemShut
  {NoStop}%
\bibitem [{\citenamefont {Frano}\ \emph {et~al.}(2020)\citenamefont {Frano},
  \citenamefont {Blanco-Canosa}, \citenamefont {Keimer},\ and\ \citenamefont
  {Birgeneau}}]{frano2020charge}%
  \BibitemOpen
  \bibfield  {author} {\bibinfo {author} {\bibfnamefont {A.}~\bibnamefont
  {Frano}}, \bibinfo {author} {\bibfnamefont {S.}~\bibnamefont
  {Blanco-Canosa}}, \bibinfo {author} {\bibfnamefont {B.}~\bibnamefont
  {Keimer}}, \ and\ \bibinfo {author} {\bibfnamefont {R.~J.}\ \bibnamefont
  {Birgeneau}},\ }\href@noop {} {\bibfield  {journal} {\bibinfo  {journal}
  {Journal of Physics: Condensed Matter}\ }\textbf {\bibinfo {volume} {32}},\
  \bibinfo {pages} {374005} (\bibinfo {year} {2020})}\BibitemShut {NoStop}%
\bibitem [{\citenamefont {Comin}\ \emph {et~al.}(2014)\citenamefont {Comin},
  \citenamefont {Frano}, \citenamefont {Yee}, \citenamefont {Yoshida},
  \citenamefont {Eisaki}, \citenamefont {Schierle}, \citenamefont {Weschke},
  \citenamefont {Sutarto}, \citenamefont {He}, \citenamefont {Soumyanarayanan},
  \citenamefont {He}, \citenamefont {Le~Tacon}, \citenamefont {Elfimov},
  \citenamefont {Hoffman}, \citenamefont {Sawatzky}, \citenamefont {Keimer},\
  and\ \citenamefont {Damascelli}}]{Comin14}%
  \BibitemOpen
  \bibfield  {author} {\bibinfo {author} {\bibfnamefont {R.}~\bibnamefont
  {Comin}}, \bibinfo {author} {\bibfnamefont {A.}~\bibnamefont {Frano}},
  \bibinfo {author} {\bibfnamefont {M.~M.}\ \bibnamefont {Yee}}, \bibinfo
  {author} {\bibfnamefont {Y.}~\bibnamefont {Yoshida}}, \bibinfo {author}
  {\bibfnamefont {H.}~\bibnamefont {Eisaki}}, \bibinfo {author} {\bibfnamefont
  {E.}~\bibnamefont {Schierle}}, \bibinfo {author} {\bibfnamefont
  {E.}~\bibnamefont {Weschke}}, \bibinfo {author} {\bibfnamefont
  {R.}~\bibnamefont {Sutarto}}, \bibinfo {author} {\bibfnamefont
  {F.}~\bibnamefont {He}}, \bibinfo {author} {\bibfnamefont {A.}~\bibnamefont
  {Soumyanarayanan}}, \bibinfo {author} {\bibfnamefont {Y.}~\bibnamefont {He}},
  \bibinfo {author} {\bibfnamefont {M.}~\bibnamefont {Le~Tacon}}, \bibinfo
  {author} {\bibfnamefont {I.~S.}\ \bibnamefont {Elfimov}}, \bibinfo {author}
  {\bibfnamefont {J.~E.}\ \bibnamefont {Hoffman}}, \bibinfo {author}
  {\bibfnamefont {G.~A.}\ \bibnamefont {Sawatzky}}, \bibinfo {author}
  {\bibfnamefont {B.}~\bibnamefont {Keimer}}, \ and\ \bibinfo {author}
  {\bibfnamefont {A.}~\bibnamefont {Damascelli}},\ }\href {\doibase
  10.1126/science.1242996} {\bibfield  {journal} {\bibinfo  {journal}
  {Science}\ }\textbf {\bibinfo {volume} {343}},\ \bibinfo {pages} {390}
  (\bibinfo {year} {2014})}\BibitemShut {NoStop}%
\bibitem [{\citenamefont {da~Silva~Neto}\ \emph {et~al.}(2014)\citenamefont
  {da~Silva~Neto}, \citenamefont {Aynajian}, \citenamefont {Frano},
  \citenamefont {Comin}, \citenamefont {Schierle}, \citenamefont {Weschke},
  \citenamefont {Gyenis}, \citenamefont {Wen}, \citenamefont {Schneeloch},
  \citenamefont {Xu}, \citenamefont {Ono}, \citenamefont {Gu}, \citenamefont
  {Le~Tacon},\ and\ \citenamefont {Yazdani}}]{daSilvaNeto14}%
  \BibitemOpen
  \bibfield  {author} {\bibinfo {author} {\bibfnamefont {E.~H.}\ \bibnamefont
  {da~Silva~Neto}}, \bibinfo {author} {\bibfnamefont {P.}~\bibnamefont
  {Aynajian}}, \bibinfo {author} {\bibfnamefont {A.}~\bibnamefont {Frano}},
  \bibinfo {author} {\bibfnamefont {R.}~\bibnamefont {Comin}}, \bibinfo
  {author} {\bibfnamefont {E.}~\bibnamefont {Schierle}}, \bibinfo {author}
  {\bibfnamefont {E.}~\bibnamefont {Weschke}}, \bibinfo {author} {\bibfnamefont
  {A.}~\bibnamefont {Gyenis}}, \bibinfo {author} {\bibfnamefont
  {J.}~\bibnamefont {Wen}}, \bibinfo {author} {\bibfnamefont {J.}~\bibnamefont
  {Schneeloch}}, \bibinfo {author} {\bibfnamefont {Z.}~\bibnamefont {Xu}},
  \bibinfo {author} {\bibfnamefont {S.}~\bibnamefont {Ono}}, \bibinfo {author}
  {\bibfnamefont {G.}~\bibnamefont {Gu}}, \bibinfo {author} {\bibfnamefont
  {M.}~\bibnamefont {Le~Tacon}}, \ and\ \bibinfo {author} {\bibfnamefont
  {A.}~\bibnamefont {Yazdani}},\ }\href {\doibase 10.1126/science.1243479}
  {\bibfield  {journal} {\bibinfo  {journal} {Science}\ }\textbf {\bibinfo
  {volume} {343}},\ \bibinfo {pages} {393} (\bibinfo {year}
  {2014})}\BibitemShut {NoStop}%
\bibitem [{\citenamefont {Fink}\ \emph {et~al.}(2009)\citenamefont {Fink},
  \citenamefont {Schierle}, \citenamefont {Weschke}, \citenamefont {Geck},
  \citenamefont {Hawthorn}, \citenamefont {Soltwisch}, \citenamefont {Wadati},
  \citenamefont {Wu}, \citenamefont {D\"urr}, \citenamefont {Wizent},
  \citenamefont {B\"uchner},\ and\ \citenamefont {Sawatzky}}]{Fink09}%
  \BibitemOpen
  \bibfield  {author} {\bibinfo {author} {\bibfnamefont {J.}~\bibnamefont
  {Fink}}, \bibinfo {author} {\bibfnamefont {E.}~\bibnamefont {Schierle}},
  \bibinfo {author} {\bibfnamefont {E.}~\bibnamefont {Weschke}}, \bibinfo
  {author} {\bibfnamefont {J.}~\bibnamefont {Geck}}, \bibinfo {author}
  {\bibfnamefont {D.}~\bibnamefont {Hawthorn}}, \bibinfo {author}
  {\bibfnamefont {V.}~\bibnamefont {Soltwisch}}, \bibinfo {author}
  {\bibfnamefont {H.}~\bibnamefont {Wadati}}, \bibinfo {author} {\bibfnamefont
  {H.-H.}\ \bibnamefont {Wu}}, \bibinfo {author} {\bibfnamefont {H.~A.}\
  \bibnamefont {D\"urr}}, \bibinfo {author} {\bibfnamefont {N.}~\bibnamefont
  {Wizent}}, \bibinfo {author} {\bibfnamefont {B.}~\bibnamefont {B\"uchner}}, \
  and\ \bibinfo {author} {\bibfnamefont {G.~A.}\ \bibnamefont {Sawatzky}},\
  }\href {\doibase 10.1103/PhysRevB.79.100502} {\bibfield  {journal} {\bibinfo
  {journal} {Phys. Rev. B}\ }\textbf {\bibinfo {volume} {79}},\ \bibinfo
  {pages} {100502(R)} (\bibinfo {year} {2009})}\BibitemShut {NoStop}%
\bibitem [{\citenamefont {Fink}\ \emph {et~al.}(2011)\citenamefont {Fink},
  \citenamefont {Soltwisch}, \citenamefont {Geck}, \citenamefont {Schierle},
  \citenamefont {Weschke},\ and\ \citenamefont {B\"uchner}}]{Fink11}%
  \BibitemOpen
  \bibfield  {author} {\bibinfo {author} {\bibfnamefont {J.}~\bibnamefont
  {Fink}}, \bibinfo {author} {\bibfnamefont {V.}~\bibnamefont {Soltwisch}},
  \bibinfo {author} {\bibfnamefont {J.}~\bibnamefont {Geck}}, \bibinfo {author}
  {\bibfnamefont {E.}~\bibnamefont {Schierle}}, \bibinfo {author}
  {\bibfnamefont {E.}~\bibnamefont {Weschke}}, \ and\ \bibinfo {author}
  {\bibfnamefont {B.}~\bibnamefont {B\"uchner}},\ }\href {\doibase
  10.1103/PhysRevB.83.092503} {\bibfield  {journal} {\bibinfo  {journal} {Phys.
  Rev. B}\ }\textbf {\bibinfo {volume} {83}},\ \bibinfo {pages} {092503}
  (\bibinfo {year} {2011})}\BibitemShut {NoStop}%
\bibitem [{\citenamefont {H\"ucker}\ \emph {et~al.}(2011)\citenamefont
  {H\"ucker}, \citenamefont {v.~Zimmermann}, \citenamefont {Gu}, \citenamefont
  {Xu}, \citenamefont {Wen}, \citenamefont {Xu}, \citenamefont {Kang},
  \citenamefont {Zheludev},\ and\ \citenamefont {Tranquada}}]{Hucker11}%
  \BibitemOpen
  \bibfield  {author} {\bibinfo {author} {\bibfnamefont {M.}~\bibnamefont
  {H\"ucker}}, \bibinfo {author} {\bibfnamefont {M.}~\bibnamefont
  {v.~Zimmermann}}, \bibinfo {author} {\bibfnamefont {G.~D.}\ \bibnamefont
  {Gu}}, \bibinfo {author} {\bibfnamefont {Z.~J.}\ \bibnamefont {Xu}}, \bibinfo
  {author} {\bibfnamefont {J.~S.}\ \bibnamefont {Wen}}, \bibinfo {author}
  {\bibfnamefont {G.}~\bibnamefont {Xu}}, \bibinfo {author} {\bibfnamefont
  {H.~J.}\ \bibnamefont {Kang}}, \bibinfo {author} {\bibfnamefont
  {A.}~\bibnamefont {Zheludev}}, \ and\ \bibinfo {author} {\bibfnamefont
  {J.~M.}\ \bibnamefont {Tranquada}},\ }\href {\doibase
  10.1103/PhysRevB.83.104506} {\bibfield  {journal} {\bibinfo  {journal} {Phys.
  Rev. B}\ }\textbf {\bibinfo {volume} {83}},\ \bibinfo {pages} {104506}
  (\bibinfo {year} {2011})}\BibitemShut {NoStop}%
\bibitem [{\citenamefont {Forgan}\ \emph {et~al.}(2015)\citenamefont {Forgan},
  \citenamefont {Blackburn}, \citenamefont {Holmes}, \citenamefont {Briffa},
  \citenamefont {Chang}, \citenamefont {Bouchenoire}, \citenamefont {Brown},
  \citenamefont {Liang}, \citenamefont {Bonn}, \citenamefont {Hardy},
  \citenamefont {Christensen}, \citenamefont {Zimmermann}, \citenamefont
  {H{\"u}cker},\ and\ \citenamefont {Hayden}}]{forgan2015}%
  \BibitemOpen
  \bibfield  {author} {\bibinfo {author} {\bibfnamefont {E.~M.}\ \bibnamefont
  {Forgan}}, \bibinfo {author} {\bibfnamefont {E.}~\bibnamefont {Blackburn}},
  \bibinfo {author} {\bibfnamefont {A.~T.}\ \bibnamefont {Holmes}}, \bibinfo
  {author} {\bibfnamefont {A.~K.~R.}\ \bibnamefont {Briffa}}, \bibinfo {author}
  {\bibfnamefont {J.}~\bibnamefont {Chang}}, \bibinfo {author} {\bibfnamefont
  {L.}~\bibnamefont {Bouchenoire}}, \bibinfo {author} {\bibfnamefont {S.~D.}\
  \bibnamefont {Brown}}, \bibinfo {author} {\bibfnamefont {R.}~\bibnamefont
  {Liang}}, \bibinfo {author} {\bibfnamefont {D.}~\bibnamefont {Bonn}},
  \bibinfo {author} {\bibfnamefont {W.~N.}\ \bibnamefont {Hardy}}, \bibinfo
  {author} {\bibfnamefont {N.~B.}\ \bibnamefont {Christensen}}, \bibinfo
  {author} {\bibfnamefont {M.~V.}\ \bibnamefont {Zimmermann}}, \bibinfo
  {author} {\bibfnamefont {M.}~\bibnamefont {H{\"u}cker}}, \ and\ \bibinfo
  {author} {\bibfnamefont {S.~M.}\ \bibnamefont {Hayden}},\ }\href {\doibase
  10.1038/ncomms10064} {\bibfield  {journal} {\bibinfo  {journal} {Nature
  Communications}\ }\textbf {\bibinfo {volume} {6}},\ \bibinfo {pages} {10064}
  (\bibinfo {year} {2015})}\BibitemShut {NoStop}%
\bibitem [{\citenamefont {Comin}\ \emph {et~al.}(2015)\citenamefont {Comin},
  \citenamefont {Sutarto}, \citenamefont {He}, \citenamefont {da~Silva~Neto},
  \citenamefont {Chauviere}, \citenamefont {Frano}, \citenamefont {Liang},
  \citenamefont {Hardy}, \citenamefont {Bonn}, \citenamefont {Yoshida},
  \citenamefont {Eisaki}, \citenamefont {Achkar}, \citenamefont {Hawthorn},
  \citenamefont {Keimer}, \citenamefont {Sawatzky},\ and\ \citenamefont
  {Damascelli}}]{Comin:2015ca}%
  \BibitemOpen
  \bibfield  {author} {\bibinfo {author} {\bibfnamefont {R.}~\bibnamefont
  {Comin}}, \bibinfo {author} {\bibfnamefont {R.}~\bibnamefont {Sutarto}},
  \bibinfo {author} {\bibfnamefont {F.}~\bibnamefont {He}}, \bibinfo {author}
  {\bibfnamefont {E.~H.}\ \bibnamefont {da~Silva~Neto}}, \bibinfo {author}
  {\bibfnamefont {L.}~\bibnamefont {Chauviere}}, \bibinfo {author}
  {\bibfnamefont {A.}~\bibnamefont {Frano}}, \bibinfo {author} {\bibfnamefont
  {R.}~\bibnamefont {Liang}}, \bibinfo {author} {\bibfnamefont {W.~N.}\
  \bibnamefont {Hardy}}, \bibinfo {author} {\bibfnamefont {D.~A.}\ \bibnamefont
  {Bonn}}, \bibinfo {author} {\bibfnamefont {Y.}~\bibnamefont {Yoshida}},
  \bibinfo {author} {\bibfnamefont {H.}~\bibnamefont {Eisaki}}, \bibinfo
  {author} {\bibfnamefont {A.~J.}\ \bibnamefont {Achkar}}, \bibinfo {author}
  {\bibfnamefont {D.~G.}\ \bibnamefont {Hawthorn}}, \bibinfo {author}
  {\bibfnamefont {B.}~\bibnamefont {Keimer}}, \bibinfo {author} {\bibfnamefont
  {G.~A.}\ \bibnamefont {Sawatzky}}, \ and\ \bibinfo {author} {\bibfnamefont
  {A.}~\bibnamefont {Damascelli}},\ }\href {\doibase 10.1038/nmat4295}
  {\bibfield  {journal} {\bibinfo  {journal} {Nature Materials}\ }\textbf
  {\bibinfo {volume} {14}},\ \bibinfo {pages} {796} (\bibinfo {year}
  {2015})}\BibitemShut {NoStop}%
\bibitem [{\citenamefont {{Comin}}\ \emph {et~al.}(2015)\citenamefont
  {{Comin}}, \citenamefont {{Sutarto}}, \citenamefont {{da Silva Neto}},
  \citenamefont {{Chauviere}}, \citenamefont {{Liang}}, \citenamefont
  {{Hardy}}, \citenamefont {{Bonn}}, \citenamefont {{He}}, \citenamefont
  {{Sawatzky}},\ and\ \citenamefont {{Damascelli}}}]{Comin15a}%
  \BibitemOpen
  \bibfield  {author} {\bibinfo {author} {\bibfnamefont {R.}~\bibnamefont
  {{Comin}}}, \bibinfo {author} {\bibfnamefont {R.}~\bibnamefont {{Sutarto}}},
  \bibinfo {author} {\bibfnamefont {E.~H.}\ \bibnamefont {{da Silva Neto}}},
  \bibinfo {author} {\bibfnamefont {L.}~\bibnamefont {{Chauviere}}}, \bibinfo
  {author} {\bibfnamefont {R.}~\bibnamefont {{Liang}}}, \bibinfo {author}
  {\bibfnamefont {W.~N.}\ \bibnamefont {{Hardy}}}, \bibinfo {author}
  {\bibfnamefont {D.~A.}\ \bibnamefont {{Bonn}}}, \bibinfo {author}
  {\bibfnamefont {F.}~\bibnamefont {{He}}}, \bibinfo {author} {\bibfnamefont
  {G.~A.}\ \bibnamefont {{Sawatzky}}}, \ and\ \bibinfo {author} {\bibfnamefont
  {A.}~\bibnamefont {{Damascelli}}},\ }\href {\doibase 10.1126/science.1258399}
  {\bibfield  {journal} {\bibinfo  {journal} {Science}\ }\textbf {\bibinfo
  {volume} {347}},\ \bibinfo {pages} {1335} (\bibinfo {year}
  {2015})}\BibitemShut {NoStop}%
\bibitem [{\citenamefont {{Campi}}\ \emph {et~al.}(2015)\citenamefont
  {{Campi}}, \citenamefont {{Bianconi}}, \citenamefont {{Poccia}},
  \citenamefont {{Bianconi}}, \citenamefont {{Barba}}, \citenamefont
  {{Arrighetti}}, \citenamefont {{Innocenti}}, \citenamefont {{Karpinski}},
  \citenamefont {{Zhigadlo}}, \citenamefont {{Kazakov}}, \citenamefont
  {{Burghammer}}, \citenamefont {{Zimmermann}}, \citenamefont {{Sprung}},\ and\
  \citenamefont {{Ricci}}}]{Campi15}%
  \BibitemOpen
  \bibfield  {author} {\bibinfo {author} {\bibfnamefont {G.}~\bibnamefont
  {{Campi}}}, \bibinfo {author} {\bibfnamefont {A.}~\bibnamefont {{Bianconi}}},
  \bibinfo {author} {\bibfnamefont {N.}~\bibnamefont {{Poccia}}}, \bibinfo
  {author} {\bibfnamefont {G.}~\bibnamefont {{Bianconi}}}, \bibinfo {author}
  {\bibfnamefont {L.}~\bibnamefont {{Barba}}}, \bibinfo {author} {\bibfnamefont
  {G.}~\bibnamefont {{Arrighetti}}}, \bibinfo {author} {\bibfnamefont
  {D.}~\bibnamefont {{Innocenti}}}, \bibinfo {author} {\bibfnamefont
  {J.}~\bibnamefont {{Karpinski}}}, \bibinfo {author} {\bibfnamefont {N.~D.}\
  \bibnamefont {{Zhigadlo}}}, \bibinfo {author} {\bibfnamefont {S.~M.}\
  \bibnamefont {{Kazakov}}}, \bibinfo {author} {\bibfnamefont {M.}~\bibnamefont
  {{Burghammer}}}, \bibinfo {author} {\bibfnamefont {M.~V.}\ \bibnamefont
  {{Zimmermann}}}, \bibinfo {author} {\bibfnamefont {M.}~\bibnamefont
  {{Sprung}}}, \ and\ \bibinfo {author} {\bibfnamefont {A.}~\bibnamefont
  {{Ricci}}},\ }\href {\doibase 10.1038/nature14987} {\bibfield  {journal}
  {\bibinfo  {journal} {\nat}\ }\textbf {\bibinfo {volume} {525}},\ \bibinfo
  {pages} {359} (\bibinfo {year} {2015})}\BibitemShut {NoStop}%
\bibitem [{\citenamefont {Kohsaka}\ \emph {et~al.}(2007)\citenamefont
  {Kohsaka}, \citenamefont {Taylor}, \citenamefont {Fujita}, \citenamefont
  {Schmidt}, \citenamefont {Lupien}, \citenamefont {Hanaguri}, \citenamefont
  {Azuma}, \citenamefont {Takano}, \citenamefont {Eisaki}, \citenamefont
  {Takagi}, \citenamefont {Uchida},\ and\ \citenamefont {Davis}}]{Kohsaka07}%
  \BibitemOpen
  \bibfield  {author} {\bibinfo {author} {\bibfnamefont {Y.}~\bibnamefont
  {Kohsaka}}, \bibinfo {author} {\bibfnamefont {C.}~\bibnamefont {Taylor}},
  \bibinfo {author} {\bibfnamefont {K.}~\bibnamefont {Fujita}}, \bibinfo
  {author} {\bibfnamefont {A.}~\bibnamefont {Schmidt}}, \bibinfo {author}
  {\bibfnamefont {C.}~\bibnamefont {Lupien}}, \bibinfo {author} {\bibfnamefont
  {T.}~\bibnamefont {Hanaguri}}, \bibinfo {author} {\bibfnamefont
  {M.}~\bibnamefont {Azuma}}, \bibinfo {author} {\bibfnamefont
  {M.}~\bibnamefont {Takano}}, \bibinfo {author} {\bibfnamefont
  {H.}~\bibnamefont {Eisaki}}, \bibinfo {author} {\bibfnamefont
  {H.}~\bibnamefont {Takagi}}, \bibinfo {author} {\bibfnamefont
  {S.}~\bibnamefont {Uchida}}, \ and\ \bibinfo {author} {\bibfnamefont {J.~C.}\
  \bibnamefont {Davis}},\ }\href {\doibase 10.1126/science.1138584} {\bibfield
  {journal} {\bibinfo  {journal} {Science}\ }\textbf {\bibinfo {volume}
  {315}},\ \bibinfo {pages} {1380} (\bibinfo {year} {2007})}\BibitemShut
  {NoStop}%
\bibitem [{\citenamefont {Parker}\ \emph {et~al.}(2010)\citenamefont {Parker},
  \citenamefont {Aynajian}, \citenamefont {da~Silva~Neto}, \citenamefont
  {Pushp}, \citenamefont {Ono}, \citenamefont {Wen}, \citenamefont {Xu},
  \citenamefont {Gu},\ and\ \citenamefont {Yazdani}}]{Parker:2010if}%
  \BibitemOpen
  \bibfield  {author} {\bibinfo {author} {\bibfnamefont {C.~V.~C.}\
  \bibnamefont {Parker}}, \bibinfo {author} {\bibfnamefont {P.~P.}\
  \bibnamefont {Aynajian}}, \bibinfo {author} {\bibfnamefont {E.~H.~E.}\
  \bibnamefont {da~Silva~Neto}}, \bibinfo {author} {\bibfnamefont {A.~A.}\
  \bibnamefont {Pushp}}, \bibinfo {author} {\bibfnamefont {S.~S.}\ \bibnamefont
  {Ono}}, \bibinfo {author} {\bibfnamefont {J.~J.}\ \bibnamefont {Wen}},
  \bibinfo {author} {\bibfnamefont {Z.~Z.}\ \bibnamefont {Xu}}, \bibinfo
  {author} {\bibfnamefont {G.~G.}\ \bibnamefont {Gu}}, \ and\ \bibinfo {author}
  {\bibfnamefont {A.~A.}\ \bibnamefont {Yazdani}},\ }\href {\doibase
  10.1038/nature09597} {\bibfield  {journal} {\bibinfo  {journal} {Nature}\
  }\textbf {\bibinfo {volume} {468}},\ \bibinfo {pages} {677} (\bibinfo {year}
  {2010})}\BibitemShut {NoStop}%
\bibitem [{\citenamefont {Chan}\ \emph {et~al.}(2016)\citenamefont {Chan},
  \citenamefont {Harrison}, \citenamefont {McDonald}, \citenamefont {Ramshaw},
  \citenamefont {Modic}, \citenamefont {Bari{\v{s}}i{\'{c}}},\ and\
  \citenamefont {Greven}}]{chan2016single}%
  \BibitemOpen
  \bibfield  {author} {\bibinfo {author} {\bibfnamefont {M.~K.}\ \bibnamefont
  {Chan}}, \bibinfo {author} {\bibfnamefont {N.}~\bibnamefont {Harrison}},
  \bibinfo {author} {\bibfnamefont {R.~D.}\ \bibnamefont {McDonald}}, \bibinfo
  {author} {\bibfnamefont {B.~J.}\ \bibnamefont {Ramshaw}}, \bibinfo {author}
  {\bibfnamefont {K.~A.}\ \bibnamefont {Modic}}, \bibinfo {author}
  {\bibfnamefont {N.}~\bibnamefont {Bari{\v{s}}i{\'{c}}}}, \ and\ \bibinfo
  {author} {\bibfnamefont {M.}~\bibnamefont {Greven}},\ }\href {\doibase
  10.1038/ncomms12244} {\bibfield  {journal} {\bibinfo  {journal} {Nature
  Communications}\ }\textbf {\bibinfo {volume} {7}},\ \bibinfo {pages} {12244}
  (\bibinfo {year} {2016})}\BibitemShut {NoStop}%
\bibitem [{\citenamefont {Hamidian}\ \emph {et~al.}(2016)\citenamefont
  {Hamidian}, \citenamefont {Edkins}, \citenamefont {Joo}, \citenamefont
  {Kostin}, \citenamefont {Eisaki}, \citenamefont {Uchida}, \citenamefont
  {Lawler}, \citenamefont {Kim}, \citenamefont {Mackenzie}, \citenamefont
  {Fujita}, \citenamefont {Lee},\ and\ \citenamefont {Davis}}]{Hamidian16}%
  \BibitemOpen
  \bibfield  {author} {\bibinfo {author} {\bibfnamefont {M.~H.}\ \bibnamefont
  {Hamidian}}, \bibinfo {author} {\bibfnamefont {S.~D.}\ \bibnamefont
  {Edkins}}, \bibinfo {author} {\bibfnamefont {S.~H.}\ \bibnamefont {Joo}},
  \bibinfo {author} {\bibfnamefont {A.}~\bibnamefont {Kostin}}, \bibinfo
  {author} {\bibfnamefont {H.}~\bibnamefont {Eisaki}}, \bibinfo {author}
  {\bibfnamefont {S.}~\bibnamefont {Uchida}}, \bibinfo {author} {\bibfnamefont
  {M.~J.}\ \bibnamefont {Lawler}}, \bibinfo {author} {\bibfnamefont {E.-A.}\
  \bibnamefont {Kim}}, \bibinfo {author} {\bibfnamefont {A.~P.}\ \bibnamefont
  {Mackenzie}}, \bibinfo {author} {\bibfnamefont {K.}~\bibnamefont {Fujita}},
  \bibinfo {author} {\bibfnamefont {J.}~\bibnamefont {Lee}}, \ and\ \bibinfo
  {author} {\bibfnamefont {J.~C.~S.}\ \bibnamefont {Davis}},\ }\href
  {https://doi.org/10.1038/nature17411} {\bibfield  {journal} {\bibinfo
  {journal} {Nature}\ }\textbf {\bibinfo {volume} {532}},\ \bibinfo {pages}
  {343} (\bibinfo {year} {2016})}\BibitemShut {NoStop}%
\bibitem [{\citenamefont {Du}\ \emph {et~al.}(2020)\citenamefont {Du},
  \citenamefont {Li}, \citenamefont {Joo}, \citenamefont {Donoway},
  \citenamefont {Lee}, \citenamefont {Davis}, \citenamefont {Gu}, \citenamefont
  {Johnson},\ and\ \citenamefont {Fujita}}]{du2020imaging}%
  \BibitemOpen
  \bibfield  {author} {\bibinfo {author} {\bibfnamefont {Z.}~\bibnamefont
  {Du}}, \bibinfo {author} {\bibfnamefont {H.}~\bibnamefont {Li}}, \bibinfo
  {author} {\bibfnamefont {S.~H.}\ \bibnamefont {Joo}}, \bibinfo {author}
  {\bibfnamefont {E.~P.}\ \bibnamefont {Donoway}}, \bibinfo {author}
  {\bibfnamefont {J.}~\bibnamefont {Lee}}, \bibinfo {author} {\bibfnamefont
  {J.~S.}\ \bibnamefont {Davis}}, \bibinfo {author} {\bibfnamefont
  {G.}~\bibnamefont {Gu}}, \bibinfo {author} {\bibfnamefont {P.~D.}\
  \bibnamefont {Johnson}}, \ and\ \bibinfo {author} {\bibfnamefont
  {K.}~\bibnamefont {Fujita}},\ }\href@noop {} {\bibfield  {journal} {\bibinfo
  {journal} {Nature}\ }\textbf {\bibinfo {volume} {580}},\ \bibinfo {pages}
  {65} (\bibinfo {year} {2020})}\BibitemShut {NoStop}%
\bibitem [{\citenamefont {Dai}\ \emph {et~al.}(2020)\citenamefont {Dai},
  \citenamefont {Senthil},\ and\ \citenamefont {Lee}}]{Dai19}%
  \BibitemOpen
  \bibfield  {author} {\bibinfo {author} {\bibfnamefont {Z.}~\bibnamefont
  {Dai}}, \bibinfo {author} {\bibfnamefont {T.}~\bibnamefont {Senthil}}, \ and\
  \bibinfo {author} {\bibfnamefont {P.~A.}\ \bibnamefont {Lee}},\ }\href
  {\doibase 10.1103/PhysRevB.101.064502} {\bibfield  {journal} {\bibinfo
  {journal} {Phys. Rev. B}\ }\textbf {\bibinfo {volume} {101}},\ \bibinfo
  {pages} {064502} (\bibinfo {year} {2020})}\BibitemShut {NoStop}%
\bibitem [{\citenamefont {Chakraborty}\ \emph {et~al.}(2019)\citenamefont
  {Chakraborty}, \citenamefont {Grandadam}, \citenamefont {Hamidian},
  \citenamefont {Davis}, \citenamefont {Sidis},\ and\ \citenamefont
  {P\'epin}}]{Chakraborty19}%
  \BibitemOpen
  \bibfield  {author} {\bibinfo {author} {\bibfnamefont {D.}~\bibnamefont
  {Chakraborty}}, \bibinfo {author} {\bibfnamefont {M.}~\bibnamefont
  {Grandadam}}, \bibinfo {author} {\bibfnamefont {M.~H.}\ \bibnamefont
  {Hamidian}}, \bibinfo {author} {\bibfnamefont {J.~C.~S.}\ \bibnamefont
  {Davis}}, \bibinfo {author} {\bibfnamefont {Y.}~\bibnamefont {Sidis}}, \ and\
  \bibinfo {author} {\bibfnamefont {C.}~\bibnamefont {P\'epin}},\ }\href
  {\doibase 10.1103/PhysRevB.100.224511} {\bibfield  {journal} {\bibinfo
  {journal} {Phys. Rev. B}\ }\textbf {\bibinfo {volume} {100}},\ \bibinfo
  {pages} {224511} (\bibinfo {year} {2019})}\BibitemShut {NoStop}%
\bibitem [{\citenamefont {Grandadam}\ \emph {et~al.}(2020)\citenamefont
  {Grandadam}, \citenamefont {Chakraborty},\ and\ \citenamefont
  {P{\'e}pin}}]{Grandadam19}%
  \BibitemOpen
  \bibfield  {author} {\bibinfo {author} {\bibfnamefont {M.}~\bibnamefont
  {Grandadam}}, \bibinfo {author} {\bibfnamefont {D.}~\bibnamefont
  {Chakraborty}}, \ and\ \bibinfo {author} {\bibfnamefont {C.}~\bibnamefont
  {P{\'e}pin}},\ }\href {\doibase 10.1007/s10948-019-05380-6} {\bibfield
  {journal} {\bibinfo  {journal} {Journal of Superconductivity and Novel
  Magnetism}\ }\textbf {\bibinfo {volume} {33}},\ \bibinfo {pages} {2361}
  (\bibinfo {year} {2020})}\BibitemShut {NoStop}%
\bibitem [{\citenamefont {Lee}(2014)}]{Lee14}%
  \BibitemOpen
  \bibfield  {author} {\bibinfo {author} {\bibfnamefont {P.~A.}\ \bibnamefont
  {Lee}},\ }\href {\doibase 10.1103/PhysRevX.4.031017} {\bibfield  {journal}
  {\bibinfo  {journal} {Phys. Rev. X}\ }\textbf {\bibinfo {volume} {4}},\
  \bibinfo {pages} {031017} (\bibinfo {year} {2014})}\BibitemShut {NoStop}%
\bibitem [{\citenamefont {Xu}\ \emph {et~al.}(2019)\citenamefont {Xu},
  \citenamefont {Law},\ and\ \citenamefont {Lee}}]{xu2019pair}%
  \BibitemOpen
  \bibfield  {author} {\bibinfo {author} {\bibfnamefont {X.~Y.}\ \bibnamefont
  {Xu}}, \bibinfo {author} {\bibfnamefont {K.~T.}\ \bibnamefont {Law}}, \ and\
  \bibinfo {author} {\bibfnamefont {P.~A.}\ \bibnamefont {Lee}},\ }\href@noop
  {} {\bibfield  {journal} {\bibinfo  {journal} {Physical review letters}\
  }\textbf {\bibinfo {volume} {122}},\ \bibinfo {pages} {167001} (\bibinfo
  {year} {2019})}\BibitemShut {NoStop}%
\bibitem [{\citenamefont {Berg}\ \emph {et~al.}(2010)\citenamefont {Berg},
  \citenamefont {Fradkin},\ and\ \citenamefont {Kivelson}}]{berg2010pair}%
  \BibitemOpen
  \bibfield  {author} {\bibinfo {author} {\bibfnamefont {E.}~\bibnamefont
  {Berg}}, \bibinfo {author} {\bibfnamefont {E.}~\bibnamefont {Fradkin}}, \
  and\ \bibinfo {author} {\bibfnamefont {S.~A.}\ \bibnamefont {Kivelson}},\
  }\href@noop {} {\bibfield  {journal} {\bibinfo  {journal} {Physical review
  letters}\ }\textbf {\bibinfo {volume} {105}},\ \bibinfo {pages} {146403}
  (\bibinfo {year} {2010})}\BibitemShut {NoStop}%
\bibitem [{\citenamefont {Peng}\ \emph {et~al.}(2021)\citenamefont {Peng},
  \citenamefont {Jiang}, \citenamefont {Devereaux},\ and\ \citenamefont
  {Jiang}}]{peng2021precursor}%
  \BibitemOpen
  \bibfield  {author} {\bibinfo {author} {\bibfnamefont {C.}~\bibnamefont
  {Peng}}, \bibinfo {author} {\bibfnamefont {Y.-F.}\ \bibnamefont {Jiang}},
  \bibinfo {author} {\bibfnamefont {T.~P.}\ \bibnamefont {Devereaux}}, \ and\
  \bibinfo {author} {\bibfnamefont {H.-C.}\ \bibnamefont {Jiang}},\ }\href
  {\doibase 10.1038/s41535-021-00363-0} {\bibfield  {journal} {\bibinfo
  {journal} {npj Quantum Materials}\ }\textbf {\bibinfo {volume} {6}},\
  \bibinfo {pages} {64} (\bibinfo {year} {2021})}\BibitemShut {NoStop}%
\bibitem [{\citenamefont {Dash}\ and\ \citenamefont
  {S{\'e}n{\'e}chal}(2021)}]{dash2021}%
  \BibitemOpen
  \bibfield  {author} {\bibinfo {author} {\bibfnamefont {S.~S.}\ \bibnamefont
  {Dash}}\ and\ \bibinfo {author} {\bibfnamefont {D.}~\bibnamefont
  {S{\'e}n{\'e}chal}},\ }\href@noop {} {\bibfield  {journal} {\bibinfo
  {journal} {Physical Review B}\ }\textbf {\bibinfo {volume} {103}},\ \bibinfo
  {pages} {045142} (\bibinfo {year} {2021})}\BibitemShut {NoStop}%
\bibitem [{\citenamefont {Sachdev}\ and\ \citenamefont
  {La~Placa}(2013)}]{Sachdev13}%
  \BibitemOpen
  \bibfield  {author} {\bibinfo {author} {\bibfnamefont {S.}~\bibnamefont
  {Sachdev}}\ and\ \bibinfo {author} {\bibfnamefont {R.}~\bibnamefont
  {La~Placa}},\ }\href {\doibase 10.1103/PhysRevLett.111.027202} {\bibfield
  {journal} {\bibinfo  {journal} {Phys. Rev. Lett.}\ }\textbf {\bibinfo
  {volume} {111}},\ \bibinfo {pages} {027202} (\bibinfo {year}
  {2013})}\BibitemShut {NoStop}%
\bibitem [{\citenamefont {Zheng}\ \emph {et~al.}(2017)\citenamefont {Zheng},
  \citenamefont {Chung}, \citenamefont {Corboz}, \citenamefont {Ehlers},
  \citenamefont {Qin}, \citenamefont {Noack}, \citenamefont {Shi},
  \citenamefont {White}, \citenamefont {Zhang},\ and\ \citenamefont
  {Chan}}]{HubbardStripe17}%
  \BibitemOpen
  \bibfield  {author} {\bibinfo {author} {\bibfnamefont {B.-X.}\ \bibnamefont
  {Zheng}}, \bibinfo {author} {\bibfnamefont {C.-M.}\ \bibnamefont {Chung}},
  \bibinfo {author} {\bibfnamefont {P.}~\bibnamefont {Corboz}}, \bibinfo
  {author} {\bibfnamefont {G.}~\bibnamefont {Ehlers}}, \bibinfo {author}
  {\bibfnamefont {M.-P.}\ \bibnamefont {Qin}}, \bibinfo {author} {\bibfnamefont
  {R.~M.}\ \bibnamefont {Noack}}, \bibinfo {author} {\bibfnamefont
  {H.}~\bibnamefont {Shi}}, \bibinfo {author} {\bibfnamefont {S.~R.}\
  \bibnamefont {White}}, \bibinfo {author} {\bibfnamefont {S.}~\bibnamefont
  {Zhang}}, \ and\ \bibinfo {author} {\bibfnamefont {G.~K.-L.}\ \bibnamefont
  {Chan}},\ }\href@noop {} {\bibfield  {journal} {\bibinfo  {journal}
  {Science}\ }\textbf {\bibinfo {volume} {358}},\ \bibinfo {pages} {1155}
  (\bibinfo {year} {2017})}\BibitemShut {NoStop}%
\bibitem [{\citenamefont {Himeda}\ \emph {et~al.}(2002)\citenamefont {Himeda},
  \citenamefont {Kato},\ and\ \citenamefont {Ogata}}]{tJPDW0}%
  \BibitemOpen
  \bibfield  {author} {\bibinfo {author} {\bibfnamefont {A.}~\bibnamefont
  {Himeda}}, \bibinfo {author} {\bibfnamefont {T.}~\bibnamefont {Kato}}, \ and\
  \bibinfo {author} {\bibfnamefont {M.}~\bibnamefont {Ogata}},\ }\href
  {\doibase 10.1103/PhysRevLett.88.117001} {\bibfield  {journal} {\bibinfo
  {journal} {Phys. Rev. Lett.}\ }\textbf {\bibinfo {volume} {88}},\ \bibinfo
  {pages} {117001} (\bibinfo {year} {2002})}\BibitemShut {NoStop}%
\bibitem [{\citenamefont {Corboz}\ \emph {et~al.}(2014)\citenamefont {Corboz},
  \citenamefont {Rice},\ and\ \citenamefont {Troyer}}]{Corboz14}%
  \BibitemOpen
  \bibfield  {author} {\bibinfo {author} {\bibfnamefont {P.}~\bibnamefont
  {Corboz}}, \bibinfo {author} {\bibfnamefont {T.~M.}\ \bibnamefont {Rice}}, \
  and\ \bibinfo {author} {\bibfnamefont {M.}~\bibnamefont {Troyer}},\ }\href
  {\doibase 10.1103/PhysRevLett.113.046402} {\bibfield  {journal} {\bibinfo
  {journal} {Phys. Rev. Lett.}\ }\textbf {\bibinfo {volume} {113}},\ \bibinfo
  {pages} {046402} (\bibinfo {year} {2014})}\BibitemShut {NoStop}%
\bibitem [{\citenamefont {Choubey}\ \emph {et~al.}(2017)\citenamefont
  {Choubey}, \citenamefont {Tu}, \citenamefont {Lee},\ and\ \citenamefont
  {Hirschfeld}}]{choubey2017}%
  \BibitemOpen
  \bibfield  {author} {\bibinfo {author} {\bibfnamefont {P.}~\bibnamefont
  {Choubey}}, \bibinfo {author} {\bibfnamefont {W.-L.}\ \bibnamefont {Tu}},
  \bibinfo {author} {\bibfnamefont {T.-K.}\ \bibnamefont {Lee}}, \ and\
  \bibinfo {author} {\bibfnamefont {P.}~\bibnamefont {Hirschfeld}},\
  }\href@noop {} {\bibfield  {journal} {\bibinfo  {journal} {New Journal of
  Physics}\ }\textbf {\bibinfo {volume} {19}},\ \bibinfo {pages} {013028}
  (\bibinfo {year} {2017})}\BibitemShut {NoStop}%
\bibitem [{\citenamefont {Choubey}\ \emph {et~al.}(2020)\citenamefont
  {Choubey}, \citenamefont {Joo}, \citenamefont {Fujita}, \citenamefont {Du},
  \citenamefont {Edkins}, \citenamefont {Hamidian}, \citenamefont {Eisaki},
  \citenamefont {Uchida}, \citenamefont {Mackenzie}, \citenamefont {Lee} \emph
  {et~al.}}]{choubey2020}%
  \BibitemOpen
  \bibfield  {author} {\bibinfo {author} {\bibfnamefont {P.}~\bibnamefont
  {Choubey}}, \bibinfo {author} {\bibfnamefont {S.~H.}\ \bibnamefont {Joo}},
  \bibinfo {author} {\bibfnamefont {K.}~\bibnamefont {Fujita}}, \bibinfo
  {author} {\bibfnamefont {Z.}~\bibnamefont {Du}}, \bibinfo {author}
  {\bibfnamefont {S.}~\bibnamefont {Edkins}}, \bibinfo {author} {\bibfnamefont
  {M.}~\bibnamefont {Hamidian}}, \bibinfo {author} {\bibfnamefont
  {H.}~\bibnamefont {Eisaki}}, \bibinfo {author} {\bibfnamefont
  {S.}~\bibnamefont {Uchida}}, \bibinfo {author} {\bibfnamefont
  {A.}~\bibnamefont {Mackenzie}}, \bibinfo {author} {\bibfnamefont
  {J.}~\bibnamefont {Lee}},  \emph {et~al.},\ }\href@noop {} {\bibfield
  {journal} {\bibinfo  {journal} {Proceedings of the National Academy of
  Sciences}\ }\textbf {\bibinfo {volume} {117}},\ \bibinfo {pages} {14805}
  (\bibinfo {year} {2020})}\BibitemShut {NoStop}%
\bibitem [{\citenamefont {Allais}\ \emph
  {et~al.}(2014{\natexlab{a}})\citenamefont {Allais}, \citenamefont {Bauer},\
  and\ \citenamefont {Sachdev}}]{Allais14c}%
  \BibitemOpen
  \bibfield  {author} {\bibinfo {author} {\bibfnamefont {A.}~\bibnamefont
  {Allais}}, \bibinfo {author} {\bibfnamefont {J.}~\bibnamefont {Bauer}}, \
  and\ \bibinfo {author} {\bibfnamefont {S.}~\bibnamefont {Sachdev}},\ }\href
  {\doibase 10.1103/PhysRevB.90.155114} {\bibfield  {journal} {\bibinfo
  {journal} {Phys. Rev. B}\ }\textbf {\bibinfo {volume} {90}},\ \bibinfo
  {pages} {155114} (\bibinfo {year} {2014}{\natexlab{a}})}\BibitemShut
  {NoStop}%
\bibitem [{\citenamefont {Allais}\ \emph
  {et~al.}(2014{\natexlab{b}})\citenamefont {Allais}, \citenamefont {Bauer},\
  and\ \citenamefont {Sachdev}}]{Allais14b}%
  \BibitemOpen
  \bibfield  {author} {\bibinfo {author} {\bibfnamefont {A.}~\bibnamefont
  {Allais}}, \bibinfo {author} {\bibfnamefont {J.}~\bibnamefont {Bauer}}, \
  and\ \bibinfo {author} {\bibfnamefont {S.}~\bibnamefont {Sachdev}},\ }\href
  {\doibase 10.1007/s12648-014-0488-4} {\bibfield  {journal} {\bibinfo
  {journal} {Indian J. Phys.}\ }\textbf {\bibinfo {volume} {88}},\ \bibinfo
  {pages} {905} (\bibinfo {year} {2014}{\natexlab{b}})}\BibitemShut {NoStop}%
\bibitem [{\citenamefont {Spa{\l}ek}\ \emph {et~al.}(2017)\citenamefont
  {Spa{\l}ek}, \citenamefont {Zegrodnik},\ and\ \citenamefont
  {Kaczmarczyk}}]{spalek2017}%
  \BibitemOpen
  \bibfield  {author} {\bibinfo {author} {\bibfnamefont {J.}~\bibnamefont
  {Spa{\l}ek}}, \bibinfo {author} {\bibfnamefont {M.}~\bibnamefont
  {Zegrodnik}}, \ and\ \bibinfo {author} {\bibfnamefont {J.}~\bibnamefont
  {Kaczmarczyk}},\ }\href@noop {} {\bibfield  {journal} {\bibinfo  {journal}
  {Physical Review B}\ }\textbf {\bibinfo {volume} {95}},\ \bibinfo {pages}
  {024506} (\bibinfo {year} {2017})}\BibitemShut {NoStop}%
\bibitem [{\citenamefont {Sau}\ and\ \citenamefont
  {Sachdev}(2014)}]{sau2014mean}%
  \BibitemOpen
  \bibfield  {author} {\bibinfo {author} {\bibfnamefont {J.~D.}\ \bibnamefont
  {Sau}}\ and\ \bibinfo {author} {\bibfnamefont {S.}~\bibnamefont {Sachdev}},\
  }\href@noop {} {\bibfield  {journal} {\bibinfo  {journal} {Physical Review
  B}\ }\textbf {\bibinfo {volume} {89}},\ \bibinfo {pages} {075129} (\bibinfo
  {year} {2014})}\BibitemShut {NoStop}%
\bibitem [{\citenamefont {Zegrodnik}\ and\ \citenamefont
  {Spa\l{}ek}(2018)}]{ZegrodnikPRB}%
  \BibitemOpen
  \bibfield  {author} {\bibinfo {author} {\bibfnamefont {M.}~\bibnamefont
  {Zegrodnik}}\ and\ \bibinfo {author} {\bibfnamefont {J.}~\bibnamefont
  {Spa\l{}ek}},\ }\href {\doibase 10.1103/PhysRevB.98.155144} {\bibfield
  {journal} {\bibinfo  {journal} {Phys. Rev. B}\ }\textbf {\bibinfo {volume}
  {98}},\ \bibinfo {pages} {155144} (\bibinfo {year} {2018})}\BibitemShut
  {NoStop}%
\bibitem [{\citenamefont {Zhao}\ \emph {et~al.}(2019)\citenamefont {Zhao},
  \citenamefont {Ren}, \citenamefont {Rachmilowitz}, \citenamefont
  {Schneeloch}, \citenamefont {Zhong}, \citenamefont {Gu}, \citenamefont
  {Wang},\ and\ \citenamefont {Zeljkovic}}]{zhao2019charge}%
  \BibitemOpen
  \bibfield  {author} {\bibinfo {author} {\bibfnamefont {H.}~\bibnamefont
  {Zhao}}, \bibinfo {author} {\bibfnamefont {Z.}~\bibnamefont {Ren}}, \bibinfo
  {author} {\bibfnamefont {B.}~\bibnamefont {Rachmilowitz}}, \bibinfo {author}
  {\bibfnamefont {J.}~\bibnamefont {Schneeloch}}, \bibinfo {author}
  {\bibfnamefont {R.}~\bibnamefont {Zhong}}, \bibinfo {author} {\bibfnamefont
  {G.}~\bibnamefont {Gu}}, \bibinfo {author} {\bibfnamefont {Z.}~\bibnamefont
  {Wang}}, \ and\ \bibinfo {author} {\bibfnamefont {I.}~\bibnamefont
  {Zeljkovic}},\ }\href@noop {} {\bibfield  {journal} {\bibinfo  {journal}
  {Nature materials}\ }\textbf {\bibinfo {volume} {18}},\ \bibinfo {pages}
  {103} (\bibinfo {year} {2019})}\BibitemShut {NoStop}%
\bibitem [{\citenamefont {Chao}\ \emph {et~al.}(1977)\citenamefont {Chao},
  \citenamefont {Spalek},\ and\ \citenamefont {Oles}}]{chao1977kinetic}%
  \BibitemOpen
  \bibfield  {author} {\bibinfo {author} {\bibfnamefont {K.}~\bibnamefont
  {Chao}}, \bibinfo {author} {\bibfnamefont {J.}~\bibnamefont {Spalek}}, \ and\
  \bibinfo {author} {\bibfnamefont {A.}~\bibnamefont {Oles}},\ }\href@noop {}
  {\bibfield  {journal} {\bibinfo  {journal} {Journal of Physics C: Solid State
  Physics}\ }\textbf {\bibinfo {volume} {10}},\ \bibinfo {pages} {L271}
  (\bibinfo {year} {1977})}\BibitemShut {NoStop}%
\bibitem [{\citenamefont {Fazekas}(1999)}]{fazekas1999lecture}%
  \BibitemOpen
  \bibfield  {author} {\bibinfo {author} {\bibfnamefont {P.}~\bibnamefont
  {Fazekas}},\ }\href@noop {} {\emph {\bibinfo {title} {Lecture notes on
  electron correlation and magnetism}}},\ Vol.~\bibinfo {volume} {5}\ (\bibinfo
   {publisher} {World scientific},\ \bibinfo {year} {1999})\BibitemShut
  {NoStop}%
\bibitem [{\citenamefont {Fukushima}(2008)}]{fukushima2008grand}%
  \BibitemOpen
  \bibfield  {author} {\bibinfo {author} {\bibfnamefont {N.}~\bibnamefont
  {Fukushima}},\ }\href@noop {} {\bibfield  {journal} {\bibinfo  {journal}
  {Physical Review B}\ }\textbf {\bibinfo {volume} {78}},\ \bibinfo {pages}
  {115105} (\bibinfo {year} {2008})}\BibitemShut {NoStop}%
\bibitem [{\citenamefont {Ko}\ \emph {et~al.}(2007)\citenamefont {Ko},
  \citenamefont {Nave},\ and\ \citenamefont {Lee}}]{ko2007extended}%
  \BibitemOpen
  \bibfield  {author} {\bibinfo {author} {\bibfnamefont {W.-H.}\ \bibnamefont
  {Ko}}, \bibinfo {author} {\bibfnamefont {C.~P.}\ \bibnamefont {Nave}}, \ and\
  \bibinfo {author} {\bibfnamefont {P.~A.}\ \bibnamefont {Lee}},\ }\href@noop
  {} {\bibfield  {journal} {\bibinfo  {journal} {Physical Review B}\ }\textbf
  {\bibinfo {volume} {76}},\ \bibinfo {pages} {245113} (\bibinfo {year}
  {2007})}\BibitemShut {NoStop}%
\bibitem [{\citenamefont {Chakraborty}\ and\ \citenamefont
  {Ghosal}(2014)}]{Chakraborty:2014iq}%
  \BibitemOpen
  \bibfield  {author} {\bibinfo {author} {\bibfnamefont {D.}~\bibnamefont
  {Chakraborty}}\ and\ \bibinfo {author} {\bibfnamefont {A.}~\bibnamefont
  {Ghosal}},\ }\href {\doibase 10.1088/1367-2630/16/10/103018} {\bibfield
  {journal} {\bibinfo  {journal} {New Journal of Physics}\ }\textbf {\bibinfo
  {volume} {16}},\ \bibinfo {pages} {103018} (\bibinfo {year}
  {2014})}\BibitemShut {NoStop}%
\bibitem [{\citenamefont {Garg}\ \emph {et~al.}(2008)\citenamefont {Garg},
  \citenamefont {Randeria},\ and\ \citenamefont {Trivedi}}]{garg2008strong}%
  \BibitemOpen
  \bibfield  {author} {\bibinfo {author} {\bibfnamefont {A.}~\bibnamefont
  {Garg}}, \bibinfo {author} {\bibfnamefont {M.}~\bibnamefont {Randeria}}, \
  and\ \bibinfo {author} {\bibfnamefont {N.}~\bibnamefont {Trivedi}},\
  }\href@noop {} {\bibfield  {journal} {\bibinfo  {journal} {Nature physics}\
  }\textbf {\bibinfo {volume} {4}},\ \bibinfo {pages} {762} (\bibinfo {year}
  {2008})}\BibitemShut {NoStop}%
\bibitem [{\citenamefont {De~Gennes}\ and\ \citenamefont
  {Pincus}(2018)}]{PdGBook}%
  \BibitemOpen
  \bibfield  {author} {\bibinfo {author} {\bibfnamefont {P.-G.}\ \bibnamefont
  {De~Gennes}}\ and\ \bibinfo {author} {\bibfnamefont {P.~A.}\ \bibnamefont
  {Pincus}},\ }\href@noop {} {\emph {\bibinfo {title} {Superconductivity of
  metals and alloys}}}\ (\bibinfo  {publisher} {CRC Press},\ \bibinfo {year}
  {2018})\BibitemShut {NoStop}%
\bibitem [{\citenamefont {Ghosal}\ \emph {et~al.}(1998)\citenamefont {Ghosal},
  \citenamefont {Randeria},\ and\ \citenamefont {Trivedi}}]{Ghosal98}%
  \BibitemOpen
  \bibfield  {author} {\bibinfo {author} {\bibfnamefont {A.}~\bibnamefont
  {Ghosal}}, \bibinfo {author} {\bibfnamefont {M.}~\bibnamefont {Randeria}}, \
  and\ \bibinfo {author} {\bibfnamefont {N.}~\bibnamefont {Trivedi}},\ }\href
  {\doibase 10.1103/PhysRevLett.81.3940} {\bibfield  {journal} {\bibinfo
  {journal} {Phys. Rev. Lett.}\ }\textbf {\bibinfo {volume} {81}},\ \bibinfo
  {pages} {3940} (\bibinfo {year} {1998})}\BibitemShut {NoStop}%
\bibitem [{\citenamefont {Ghosal}\ \emph {et~al.}(2001)\citenamefont {Ghosal},
  \citenamefont {Randeria},\ and\ \citenamefont {Trivedi}}]{Ghosal01}%
  \BibitemOpen
  \bibfield  {author} {\bibinfo {author} {\bibfnamefont {A.}~\bibnamefont
  {Ghosal}}, \bibinfo {author} {\bibfnamefont {M.}~\bibnamefont {Randeria}}, \
  and\ \bibinfo {author} {\bibfnamefont {N.}~\bibnamefont {Trivedi}},\ }\href
  {\doibase 10.1103/PhysRevB.65.014501} {\bibfield  {journal} {\bibinfo
  {journal} {Phys. Rev. B}\ }\textbf {\bibinfo {volume} {65}},\ \bibinfo
  {pages} {014501} (\bibinfo {year} {2001})}\BibitemShut {NoStop}%
\bibitem [{\citenamefont {Vershinin}(2004)}]{Vershinin:2004gk}%
  \BibitemOpen
  \bibfield  {author} {\bibinfo {author} {\bibfnamefont {M.}~\bibnamefont
  {Vershinin}},\ }\href {\doibase 10.1126/science.1093384} {\bibfield
  {journal} {\bibinfo  {journal} {Science}\ }\textbf {\bibinfo {volume}
  {303}},\ \bibinfo {pages} {1995} (\bibinfo {year} {2004})}\BibitemShut
  {NoStop}%
\bibitem [{\citenamefont {El~Baggari}\ \emph {et~al.}(2018)\citenamefont
  {El~Baggari}, \citenamefont {Savitzky}, \citenamefont {Admasu}, \citenamefont
  {Kim}, \citenamefont {Cheong}, \citenamefont {Hovden},\ and\ \citenamefont
  {Kourkoutis}}]{el2018nature}%
  \BibitemOpen
  \bibfield  {author} {\bibinfo {author} {\bibfnamefont {I.}~\bibnamefont
  {El~Baggari}}, \bibinfo {author} {\bibfnamefont {B.~H.}\ \bibnamefont
  {Savitzky}}, \bibinfo {author} {\bibfnamefont {A.~S.}\ \bibnamefont
  {Admasu}}, \bibinfo {author} {\bibfnamefont {J.}~\bibnamefont {Kim}},
  \bibinfo {author} {\bibfnamefont {S.-W.}\ \bibnamefont {Cheong}}, \bibinfo
  {author} {\bibfnamefont {R.}~\bibnamefont {Hovden}}, \ and\ \bibinfo {author}
  {\bibfnamefont {L.~F.}\ \bibnamefont {Kourkoutis}},\ }\href@noop {}
  {\bibfield  {journal} {\bibinfo  {journal} {Proceedings of the National
  Academy of Sciences}\ }\textbf {\bibinfo {volume} {115}},\ \bibinfo {pages}
  {1445} (\bibinfo {year} {2018})}\BibitemShut {NoStop}%
\bibitem [{Note1()}]{Note1}%
  \BibitemOpen
  \bibinfo {note} {Note that the presence of the weak nematicity generates a
  small value of mean $\Delta _s$.}\BibitemShut {Stop}%
\bibitem [{\citenamefont {Tu}\ and\ \citenamefont {Lee}(2019)}]{tu2019}%
  \BibitemOpen
  \bibfield  {author} {\bibinfo {author} {\bibfnamefont {W.-L.}\ \bibnamefont
  {Tu}}\ and\ \bibinfo {author} {\bibfnamefont {T.-K.}\ \bibnamefont {Lee}},\
  }\href {\doibase 10.1038/s41598-018-38288-7} {\bibfield  {journal} {\bibinfo
  {journal} {Scientific Reports}\ }\textbf {\bibinfo {volume} {9}},\ \bibinfo
  {pages} {1719} (\bibinfo {year} {2019})}\BibitemShut {NoStop}%
\bibitem [{\citenamefont {Nie}\ \emph {et~al.}(2017)\citenamefont {Nie},
  \citenamefont {Maharaj}, \citenamefont {Fradkin},\ and\ \citenamefont
  {Kivelson}}]{VestigeNie}%
  \BibitemOpen
  \bibfield  {author} {\bibinfo {author} {\bibfnamefont {L.}~\bibnamefont
  {Nie}}, \bibinfo {author} {\bibfnamefont {A.~V.}\ \bibnamefont {Maharaj}},
  \bibinfo {author} {\bibfnamefont {E.}~\bibnamefont {Fradkin}}, \ and\
  \bibinfo {author} {\bibfnamefont {S.~A.}\ \bibnamefont {Kivelson}},\ }\href
  {\doibase 10.1103/PhysRevB.96.085142} {\bibfield  {journal} {\bibinfo
  {journal} {Phys. Rev. B}\ }\textbf {\bibinfo {volume} {96}},\ \bibinfo
  {pages} {085142} (\bibinfo {year} {2017})}\BibitemShut {NoStop}%
\bibitem [{\citenamefont {Miao}\ \emph {et~al.}(2017)\citenamefont {Miao},
  \citenamefont {Lorenzana}, \citenamefont {Seibold}, \citenamefont {Peng},
  \citenamefont {Amorese}, \citenamefont {Yakhou-Harris}, \citenamefont
  {Kummer}, \citenamefont {Brookes}, \citenamefont {Konik}, \citenamefont
  {Thampy} \emph {et~al.}}]{miao17}%
  \BibitemOpen
  \bibfield  {author} {\bibinfo {author} {\bibfnamefont {H.}~\bibnamefont
  {Miao}}, \bibinfo {author} {\bibfnamefont {J.}~\bibnamefont {Lorenzana}},
  \bibinfo {author} {\bibfnamefont {G.}~\bibnamefont {Seibold}}, \bibinfo
  {author} {\bibfnamefont {Y.}~\bibnamefont {Peng}}, \bibinfo {author}
  {\bibfnamefont {A.}~\bibnamefont {Amorese}}, \bibinfo {author} {\bibfnamefont
  {F.}~\bibnamefont {Yakhou-Harris}}, \bibinfo {author} {\bibfnamefont
  {K.}~\bibnamefont {Kummer}}, \bibinfo {author} {\bibfnamefont
  {N.}~\bibnamefont {Brookes}}, \bibinfo {author} {\bibfnamefont
  {R.}~\bibnamefont {Konik}}, \bibinfo {author} {\bibfnamefont
  {V.}~\bibnamefont {Thampy}},  \emph {et~al.},\ }\href@noop {} {\bibfield
  {journal} {\bibinfo  {journal} {Proceedings of the National Academy of
  Sciences}\ }\textbf {\bibinfo {volume} {114}},\ \bibinfo {pages} {12430}
  (\bibinfo {year} {2017})}\BibitemShut {NoStop}%
\bibitem [{\citenamefont {Reymbaut}\ \emph {et~al.}(2016)\citenamefont
  {Reymbaut}, \citenamefont {Charlebois}, \citenamefont {Asiani}, \citenamefont
  {Fratino}, \citenamefont {S{\'e}mon}, \citenamefont {Sordi},\ and\
  \citenamefont {Tremblay}}]{VtermImp0}%
  \BibitemOpen
  \bibfield  {author} {\bibinfo {author} {\bibfnamefont {A.}~\bibnamefont
  {Reymbaut}}, \bibinfo {author} {\bibfnamefont {M.}~\bibnamefont
  {Charlebois}}, \bibinfo {author} {\bibfnamefont {M.~F.}\ \bibnamefont
  {Asiani}}, \bibinfo {author} {\bibfnamefont {L.}~\bibnamefont {Fratino}},
  \bibinfo {author} {\bibfnamefont {P.}~\bibnamefont {S{\'e}mon}}, \bibinfo
  {author} {\bibfnamefont {G.}~\bibnamefont {Sordi}}, \ and\ \bibinfo {author}
  {\bibfnamefont {A.-M.}\ \bibnamefont {Tremblay}},\ }\href@noop {} {\bibfield
  {journal} {\bibinfo  {journal} {Physical Review B}\ }\textbf {\bibinfo
  {volume} {94}},\ \bibinfo {pages} {155146} (\bibinfo {year}
  {2016})}\BibitemShut {NoStop}%
\bibitem [{\citenamefont {S\'en\'echal}\ \emph {et~al.}(2013)\citenamefont
  {S\'en\'echal}, \citenamefont {Day}, \citenamefont {Bouliane},\ and\
  \citenamefont {Tremblay}}]{VtermImp1}%
  \BibitemOpen
  \bibfield  {author} {\bibinfo {author} {\bibfnamefont {D.}~\bibnamefont
  {S\'en\'echal}}, \bibinfo {author} {\bibfnamefont {A.~G.~R.}\ \bibnamefont
  {Day}}, \bibinfo {author} {\bibfnamefont {V.}~\bibnamefont {Bouliane}}, \
  and\ \bibinfo {author} {\bibfnamefont {A.-M.~S.}\ \bibnamefont {Tremblay}},\
  }\href {\doibase 10.1103/PhysRevB.87.075123} {\bibfield  {journal} {\bibinfo
  {journal} {Phys. Rev. B}\ }\textbf {\bibinfo {volume} {87}},\ \bibinfo
  {pages} {075123} (\bibinfo {year} {2013})}\BibitemShut {NoStop}%
\bibitem [{\citenamefont {Jiang}\ \emph {et~al.}(2018)\citenamefont {Jiang},
  \citenamefont {H{\"a}hner}, \citenamefont {Schulthess},\ and\ \citenamefont
  {Maier}}]{VtermImp2}%
  \BibitemOpen
  \bibfield  {author} {\bibinfo {author} {\bibfnamefont {M.}~\bibnamefont
  {Jiang}}, \bibinfo {author} {\bibfnamefont {U.~R.}\ \bibnamefont
  {H{\"a}hner}}, \bibinfo {author} {\bibfnamefont {T.~C.}\ \bibnamefont
  {Schulthess}}, \ and\ \bibinfo {author} {\bibfnamefont {T.~A.}\ \bibnamefont
  {Maier}},\ }\href@noop {} {\bibfield  {journal} {\bibinfo  {journal}
  {Physical Review B}\ }\textbf {\bibinfo {volume} {97}},\ \bibinfo {pages}
  {184507} (\bibinfo {year} {2018})}\BibitemShut {NoStop}%
\bibitem [{\citenamefont {Yang}\ \emph {et~al.}(2009)\citenamefont {Yang},
  \citenamefont {Chen}, \citenamefont {Rice}, \citenamefont {Sigrist},\ and\
  \citenamefont {Zhang}}]{yang2009nature}%
  \BibitemOpen
  \bibfield  {author} {\bibinfo {author} {\bibfnamefont {K.-Y.}\ \bibnamefont
  {Yang}}, \bibinfo {author} {\bibfnamefont {W.~Q.}\ \bibnamefont {Chen}},
  \bibinfo {author} {\bibfnamefont {T.~M.}\ \bibnamefont {Rice}}, \bibinfo
  {author} {\bibfnamefont {M.}~\bibnamefont {Sigrist}}, \ and\ \bibinfo
  {author} {\bibfnamefont {F.-C.}\ \bibnamefont {Zhang}},\ }\href@noop {}
  {\bibfield  {journal} {\bibinfo  {journal} {New Journal of Physics}\ }\textbf
  {\bibinfo {volume} {11}},\ \bibinfo {pages} {055053} (\bibinfo {year}
  {2009})}\BibitemShut {NoStop}%
\bibitem [{\citenamefont {H\"ucker}\ \emph {et~al.}(2014)\citenamefont
  {H\"ucker}, \citenamefont {Christensen}, \citenamefont {Holmes},
  \citenamefont {Blackburn}, \citenamefont {Forgan}, \citenamefont {Liang},
  \citenamefont {Bonn}, \citenamefont {Hardy}, \citenamefont {Gutowski},
  \citenamefont {Zimmermann}, \citenamefont {Hayden},\ and\ \citenamefont
  {Chang}}]{Huecker14a}%
  \BibitemOpen
  \bibfield  {author} {\bibinfo {author} {\bibfnamefont {M.}~\bibnamefont
  {H\"ucker}}, \bibinfo {author} {\bibfnamefont {N.~B.}\ \bibnamefont
  {Christensen}}, \bibinfo {author} {\bibfnamefont {A.~T.}\ \bibnamefont
  {Holmes}}, \bibinfo {author} {\bibfnamefont {E.}~\bibnamefont {Blackburn}},
  \bibinfo {author} {\bibfnamefont {E.~M.}\ \bibnamefont {Forgan}}, \bibinfo
  {author} {\bibfnamefont {R.}~\bibnamefont {Liang}}, \bibinfo {author}
  {\bibfnamefont {D.~A.}\ \bibnamefont {Bonn}}, \bibinfo {author}
  {\bibfnamefont {W.~N.}\ \bibnamefont {Hardy}}, \bibinfo {author}
  {\bibfnamefont {O.}~\bibnamefont {Gutowski}}, \bibinfo {author}
  {\bibfnamefont {M.~v.}\ \bibnamefont {Zimmermann}}, \bibinfo {author}
  {\bibfnamefont {S.~M.}\ \bibnamefont {Hayden}}, \ and\ \bibinfo {author}
  {\bibfnamefont {J.}~\bibnamefont {Chang}},\ }\href {\doibase
  10.1103/PhysRevB.90.054514} {\bibfield  {journal} {\bibinfo  {journal} {Phys.
  Rev. B}\ }\textbf {\bibinfo {volume} {90}},\ \bibinfo {pages} {054514}
  (\bibinfo {year} {2014})}\BibitemShut {NoStop}%
\bibitem [{\citenamefont {Blanco-Canosa}\ \emph {et~al.}(2014)\citenamefont
  {Blanco-Canosa}, \citenamefont {Frano}, \citenamefont {Schierle},
  \citenamefont {Porras}, \citenamefont {Loew}, \citenamefont {Minola},
  \citenamefont {Bluschke}, \citenamefont {Weschke}, \citenamefont {Keimer},\
  and\ \citenamefont {Le~Tacon}}]{Blanco-Canosa14}%
  \BibitemOpen
  \bibfield  {author} {\bibinfo {author} {\bibfnamefont {S.}~\bibnamefont
  {Blanco-Canosa}}, \bibinfo {author} {\bibfnamefont {A.}~\bibnamefont
  {Frano}}, \bibinfo {author} {\bibfnamefont {E.}~\bibnamefont {Schierle}},
  \bibinfo {author} {\bibfnamefont {J.}~\bibnamefont {Porras}}, \bibinfo
  {author} {\bibfnamefont {T.}~\bibnamefont {Loew}}, \bibinfo {author}
  {\bibfnamefont {M.}~\bibnamefont {Minola}}, \bibinfo {author} {\bibfnamefont
  {M.}~\bibnamefont {Bluschke}}, \bibinfo {author} {\bibfnamefont
  {E.}~\bibnamefont {Weschke}}, \bibinfo {author} {\bibfnamefont
  {B.}~\bibnamefont {Keimer}}, \ and\ \bibinfo {author} {\bibfnamefont
  {M.}~\bibnamefont {Le~Tacon}},\ }\href {\doibase 10.1103/PhysRevB.90.054513}
  {\bibfield  {journal} {\bibinfo  {journal} {Phys. Rev. B}\ }\textbf {\bibinfo
  {volume} {90}},\ \bibinfo {pages} {054513} (\bibinfo {year}
  {2014})}\BibitemShut {NoStop}%
\bibitem [{\citenamefont {Loret}\ \emph {et~al.}(2019)\citenamefont {Loret},
  \citenamefont {Auvray}, \citenamefont {Gallais}, \citenamefont {Cazayous},
  \citenamefont {Forget}, \citenamefont {Colson}, \citenamefont {Julien},
  \citenamefont {Paul}, \citenamefont {Civelli},\ and\ \citenamefont
  {Sacuto}}]{Loret19}%
  \BibitemOpen
  \bibfield  {author} {\bibinfo {author} {\bibfnamefont {B.}~\bibnamefont
  {Loret}}, \bibinfo {author} {\bibfnamefont {N.}~\bibnamefont {Auvray}},
  \bibinfo {author} {\bibfnamefont {Y.}~\bibnamefont {Gallais}}, \bibinfo
  {author} {\bibfnamefont {M.}~\bibnamefont {Cazayous}}, \bibinfo {author}
  {\bibfnamefont {A.}~\bibnamefont {Forget}}, \bibinfo {author} {\bibfnamefont
  {D.}~\bibnamefont {Colson}}, \bibinfo {author} {\bibfnamefont {M.-H.}\
  \bibnamefont {Julien}}, \bibinfo {author} {\bibfnamefont {I.}~\bibnamefont
  {Paul}}, \bibinfo {author} {\bibfnamefont {M.}~\bibnamefont {Civelli}}, \
  and\ \bibinfo {author} {\bibfnamefont {A.}~\bibnamefont {Sacuto}},\ }\href
  {\doibase 10.1038/s41567-019-0509-5} {\bibfield  {journal} {\bibinfo
  {journal} {Nature Physics}\ }\textbf {\bibinfo {volume} {15}},\ \bibinfo
  {pages} {771} (\bibinfo {year} {2019})}\BibitemShut {NoStop}%
\bibitem [{\citenamefont {Lee}\ \emph {et~al.}(2021)\citenamefont {Lee},
  \citenamefont {Collini}, \citenamefont {Sun}, \citenamefont {Mitrano},
  \citenamefont {Guo}, \citenamefont {Eckberg}, \citenamefont {Paglione},
  \citenamefont {Fradkin},\ and\ \citenamefont {Abbamonte}}]{Lee21}%
  \BibitemOpen
  \bibfield  {author} {\bibinfo {author} {\bibfnamefont {S.}~\bibnamefont
  {Lee}}, \bibinfo {author} {\bibfnamefont {J.}~\bibnamefont {Collini}},
  \bibinfo {author} {\bibfnamefont {S.~X.-L.}\ \bibnamefont {Sun}}, \bibinfo
  {author} {\bibfnamefont {M.}~\bibnamefont {Mitrano}}, \bibinfo {author}
  {\bibfnamefont {X.}~\bibnamefont {Guo}}, \bibinfo {author} {\bibfnamefont
  {C.}~\bibnamefont {Eckberg}}, \bibinfo {author} {\bibfnamefont
  {J.}~\bibnamefont {Paglione}}, \bibinfo {author} {\bibfnamefont
  {E.}~\bibnamefont {Fradkin}}, \ and\ \bibinfo {author} {\bibfnamefont
  {P.}~\bibnamefont {Abbamonte}},\ }\href {\doibase
  10.1103/PhysRevLett.127.027602} {\bibfield  {journal} {\bibinfo  {journal}
  {Phys. Rev. Lett.}\ }\textbf {\bibinfo {volume} {127}},\ \bibinfo {pages}
  {027602} (\bibinfo {year} {2021})}\BibitemShut {NoStop}%
\bibitem [{\citenamefont {Banerjee}\ \emph {et~al.}(2018)\citenamefont
  {Banerjee}, \citenamefont {Garg},\ and\ \citenamefont {Ghosal}}]{Banerjee18}%
  \BibitemOpen
  \bibfield  {author} {\bibinfo {author} {\bibfnamefont {A.}~\bibnamefont
  {Banerjee}}, \bibinfo {author} {\bibfnamefont {A.}~\bibnamefont {Garg}}, \
  and\ \bibinfo {author} {\bibfnamefont {A.}~\bibnamefont {Ghosal}},\ }\href
  {\doibase 10.1103/PhysRevB.98.104206} {\bibfield  {journal} {\bibinfo
  {journal} {Phys. Rev. B}\ }\textbf {\bibinfo {volume} {98}},\ \bibinfo
  {pages} {104206} (\bibinfo {year} {2018})}\BibitemShut {NoStop}%
\bibitem [{\citenamefont {Agterberg}\ and\ \citenamefont
  {Tsunetsugu}(2008)}]{ag08dislocations}%
  \BibitemOpen
  \bibfield  {author} {\bibinfo {author} {\bibfnamefont {D.}~\bibnamefont
  {Agterberg}}\ and\ \bibinfo {author} {\bibfnamefont {H.}~\bibnamefont
  {Tsunetsugu}},\ }\href@noop {} {\bibfield  {journal} {\bibinfo  {journal}
  {Nature Physics}\ }\textbf {\bibinfo {volume} {4}},\ \bibinfo {pages} {639}
  (\bibinfo {year} {2008})}\BibitemShut {NoStop}%
\end{thebibliography}%
\end{document}